\documentclass[prd,superscriptaddress,nofootinbib,floatfix]{revtex4-1}

\pdfoutput=1   
 
\usepackage{caption}
\usepackage[T1]{fontenc} % if needed
\usepackage[makeroom]{cancel}
\usepackage{physics}
\usepackage{empheq}
\usepackage{graphicx}
\usepackage{subcaption}
\usepackage{tabularx}
\usepackage{natbib}

\usepackage{amsmath}
\usepackage{soul}
\usepackage{amsfonts}
\usepackage{mathtools}
\usepackage{amssymb}
\usepackage{enumerate}
\usepackage{xcolor}
\usepackage{bm}
\usepackage{mathrsfs}
\usepackage{epstopdf}
\usepackage{url}
\usepackage{footnote}
\usepackage{textcomp}
\usepackage{dsfont}
\usepackage{hyperref}

\usepackage{enumerate}   
\usepackage{appendix}
\usepackage{textcomp}

\newcommand{\ba}{\begin{eqnarray}}
\newcommand{\ea}{\end{eqnarray}}
\newcommand{\nn}{\nonumber}
\def \be {\begin{equation}}
\def \ee {\end{equation}}
\def \bea {\begin{eqnarray}}
\def \eea {\end{eqnarray}}

\newcommand{\eq}[1]{(\ref{#1})}

\begin{document}
\title{\boldmath Resonant Islands of Effective-One-Body Dynamics}

\author{Che-Yu Chen,}
\email{b97202056@gmail.com}
\affiliation{Institute of Physics, Academia Sinica, Taipei 11529, Taiwan}

\author{Feng-Li Lin}
\email{fengli.lin@gmail.com}
\affiliation{Center of Astronomy and Gravitation, National Taiwan Normal University, Taipei 11677, Taiwan}
\affiliation{Department of Physics, National Taiwan Normal University, Taipei 11677, Taiwan}

\author{Avani Patel}
\email{avani.physics@gmail.com}
\affiliation{Center of Astronomy and Gravitation, National Taiwan Normal University, Taipei 11677, Taiwan}
\affiliation{Department of Physics, National Taiwan Normal University, Taipei 11677, Taiwan}

\date{\today}

\begin{abstract}
We study the chaotic signatures of the geodesic dynamics of a non-spinning test particle in the effective-one-body (EOB) formalism for the inspiral process of spinning binary black holes. We first show that the second order post-Newtonian (2PN) EOB dynamics is non-integrable by demonstrating that the EOB metric does not satisfy the criterion for the existence of Carter constant. We then employ the numerical study to find the plateaus of the rotation curve, which are associated with the existence of Birkhoff islands in the Poincar\'e surface of section, signifying the chaotic dynamics in the system. Our results show the signatures of chaos for the EOB dynamics, especially in the regime of interest for which the Kerr bounds of the component black holes hold. We also find that chaotic behavior is more obvious as the spin parameter $a$ of the deformed EOB background metric increases. Our results can help to uncover the implications of dynamical chaos in gravitational wave astronomy. Finally, we also present some preliminary results due to corrections at 3PN order. 
\end{abstract}

\maketitle
\tableofcontents

\section{Introduction}
 
The inspiral binary black holes are the most common sources of gravitational waves detected by LIGO/Virgo \cite{LIGOScientific:2021djp}. To detect such signals, precise theoretical gravitational waveforms should be prepared to perform the matched filtering procedure for combating the background noise in the detector. Therefore, studying the inspiral dynamics and the subsequent merger is important to obtain the gravitational waveforms. During the inspiral phase, the post-Newtonian approximation can be adopted so that the non-linear effects of Einstein's gravity can be taken into account perturbatively. However, strong gravity will be involved in the merger phase so that only numerical relativity can yield precise dynamics and waveforms. After the merger, the dynamics can be approximated by black hole perturbation theory to uncover the quasi-normal modes in the ringdown phase. Although numerical relativity can be used as a first-principle method to calculate the dynamics and waveforms for the whole coalescence of binary black holes, it is computationally very costly. Therefore, some approximate hybrid methods have been proposed to balance efficiency and accuracy requirements. Among them, the most synergistic is the effective-one-body (EOB) method \cite{Buonanno:1998gg,Damour:2000we,Damour:2001tu,Barausse:2009xi,Barausse:2011ys}. The basic idea of the EOB method is to map the binary inspiral post-Newtonian dynamics into the reduced dynamics of a test-body moving in some blackhole-like background metric. The EOB map is a generalization of the Newtonian reduction of the two-body problem. The advantage of this reduction is to give the leverage to extrapolate to merger dynamics provided by numerical relativity, and then connect to the ringdown phase which is by itself a ``one-body'' problem. In this way, the EOB method provides a systematic framework for studying the dynamics of the whole binary coalescence process with the minimal usage of numerical relativity for efficiency.

Due to the non-linearity of Einstein's gravity, the post-Newtonian (PN) Hamiltonian contains non-Coulombic higher-order interactions. Although the Newtonian two-body dynamics is integrable, the non-Coulombic potentials of the non-linear origin naturally raise the question of the integrability of the PN inspiral dynamics. This issue may significantly challenge the matched filtering techniques \cite{Cornish:2001jy} and has been addressed in the past decades. However, due to the complication of the PN potentials and the associated numerical calculations, the answer and the associated physical implication to gravitational wave observations are not always affirmative \cite{Levin:1999zx,Levin:2000md,Hughes:2000nzm,Cornish:2003ig,Schnittman:2001mz,Cornish:2002gz,Hartl:2004xr,Wu:2007zze,Wu:2010mv,2011GReGr..43.2185W}. Despite that, it is known that up to 1.5PN order (i.e., up to $(v/c)^3$ terms in PN Hamiltonian), the inspiral dynamics of binary spinning black holes is integrable \cite{Damour:2001tu,Cho:2019brd,Tanay:2020gfb,Tanay:2021bff,Morras:2021atg}. If the binary inspiral dynamics at higher PN orders is non-integrable, e.g. in the presence of spin-spin interactions \cite{Gopakumar:2005zz,Konigsdorffer:2005sc,Huang:2014ska,Wu:2015cqa,Huang:2016vfk}, then it is natural to ask how to study and characterize the chaotic behaviors arising from the non-integrability, and how they will affect the corresponding gravitational waveforms.  

The formal way of checking the integrability is to construct the commuting conserved charges in the context of symplectic dynamics \cite{arnold1989mathematical}. If the number of such conserved charges matches the number of the dynamical degrees of freedom, then the dynamics is integrable. In practical terms, the construction is involved and may need intricate treatments, such as introducing appropriate action-angle variables, to demonstrate the (non)-integrability; see for examples \cite{Tanay:2020gfb,Tanay:2021bff}. This procedure is in contrast to the construction of conserved charges of a test-body moving around the background black hole metric, for which more systematic tools exist, such as Killing-Yano tensor \cite{Frolov:2017kze,Papadopoulos:2018nvd,Compere:2021kjz}. On the other hand, if the dynamics turns out to be non-integrable, one can study the chaotic behavior by numerically evaluating the Lyapunov exponent \cite{2002ocda.book.....C}, which measures the asymptotic rate of exponential separation of nearby trajectories in phase space during the time evolution. When including black hole spins, the higher number of dimensions of phase space will sometimes cause numerical ambiguity in identifying the chaotic behaviors \cite{Levin:1999zx,Schnittman:2001mz,Cornish:2002gz,Cornish:2003ig,Hartl:2004xr,Wu:2007zze,Wu:2010mv}, see \cite{Lukes-Gerakopoulos:2016bup,Harms:2016ctx,Lukes-Gerakopoulos:2016udm,Lukes-Gerakopoulos:2017vkj,Zelenka:2019nyp,Skoupy:2021asz,Timogiannis:2021ung,Skoupy:2022adh,Timogiannis:2022bks} for more recent discussions. This is again in contrast to the numerical characterization of chaotic behavior of a test-body around a background black hole, for which the phase space is half so that the numerical evaluation takes less effort \cite{Suzuki:1996gm,Hartl:2002ig,Hartl:2003da,Kiuchi:2004bv,Gair:2007kr,Han:2008zzf,Flanagan:2010cd}. In fact, for the test-body case, one can reduce the phase space to 3-dimensional one{\footnote{After imposing the conservation of energy and angular momentum, the phase space variables contain the radial variable, the polar angle, and their respective conjugate momenta. With the constraint equation from the geodesic Hamiltonian for the constant mass of the test-body, the phase space is 3-dimensional.}} by adopting the energy and angular momentum conservation so that one can resort to the Birkhoff islands on Poincar\'e surfaces of section \cite{Apostolatos:2009vu,Lukes-Gerakopoulos:2010ipp,Contopoulos:2011dz,Brink:2013nna,Brink:2015roa,Cardenas-Avendano:2018ocb,Destounis:2020kss,Lukes-Gerakopoulos:2021ybx,Destounis:2021mqv,Lukes-Gerakopoulos:2012qpc,Zelenka:2017aqn,Lukes-Gerakopoulos:2017jub,Destounis:2021rko,Mukherjee:2022dju} to identify chaos.

By construction, the EOB dynamics is equivalent to the PN dynamics at the corresponding PN order. However, as the name of EOB suggests, it is usually assumed that the EOB dynamics is effectively one-body one, i.e., it is a good approximation to freeze the dynamics of the background metric such as mass, spin, or position of the deformed black hole. Under such approximation, the number of degrees of freedom of the EOB dynamics is almost half of the PN ones, and the remaining degrees of freedom are just the ones of the test-body. This reduction, then, will help in either formal discussion or numerical implementation of checking the integrability or characterizing the chaotic behaviors. In summary, the study of the integrability of EOB dynamics will help to reveal the approximate integrability of the corresponding PN dynamics. This approximate integrability is valid whenever the EOB dynamics with a frozen background metric is a good approximation to the original PN ones. Especially, the phase space of EOB can be reduced to 3-dimensional so that one can adopt the Poincar\'e surfaces of section and the rotation numbers to identify the chaotic behaviors. As we will demonstrate in this paper, these methods turn out to be suitable to identify chaotic behaviors of a weakly non-integrable dynamical system, such as the dynamics of a binary system discussed throughout this paper, which can be treated as a deformed integrable system according to the Kolmogorov–Arnold–Moser (KAM) theorem \cite{schuster_just_2005}.

This paper is outlined as follows. In section \ref{EOB_rev}, we briefly review the EOB dynamics for spinning binaries. This section is supplemented by Appendix \ref{app_A} in which the EOB mapping is reviewed in more detail. In section \ref{sec.pk}, we show that the EOB metric does not satisfy the criterion for the existence of Carter constant, which may imply that the geodesic dynamics of EOB metric is non-integrable. In section \ref{sec.island}, we identify the signatures of the chaos of the 2PN EOB dynamics by investigating the Poincar\'e surfaces of section and the rotation numbers of phase orbits. Possible signatures of chaos in 3PN EOB dynamics will also be discussed. We finally conclude in section \ref{sec.conclude}.

\section{A brief review of EOB dynamics}\label{EOB_rev}
The basic idea of EOB formalism is to map the two-body PN dynamics into probe dynamics moving in an ``effective metric'' by matching their corresponding reduced Hamilton-Jacobi functionals, which are expressed as a series of constants of motion. There are several ways to obtain the EOB dynamics, the canonical one was formulated \cite{Buonanno:1998gg} for the 2PN order of non-spinning binary black holes and later generalized to higher PN cases of spinning binaries, e.g., \cite{Barausse:2009aa, Barausse:2009xi, Barausse:2011ys}. In particular, the EOB method adopts a particular way of relating the total energy of the two-body system $E_{\rm EOB}$ to the effective one-body energy $E_{\rm eff}$ implied by relativistic kinematics, 
\be
E_{\rm eff}=\gamma \mu \qquad \mbox{with} \qquad   \gamma={E_{\rm EOB}^2-m_1^2-m_2^2 \over 2 m_1 m_2}\,, 
\ee
or equivalently
\be \label{E_EOB}
E_{\rm EOB}=M\sqrt{1+ {2\eta \over \mu} (E_{\rm eff}-1)}
\ee
where $m_{1,2}$ are the masses of two black holes. The total mass, reduced mass, and the symmetric mass ratio are, respectively, given by
\be
M=m_1+ m_2\,, \qquad \mu={m_1m_2 \over M}\,, \qquad \eta={\mu \over M}\,,
\ee
and $\gamma=1/\sqrt{1-\vec{v}^2}$ is the Lorentz contraction factor of the relative velocity $\vec{v}$. Note that the symmetric mass ratio $\eta$ ranges from $0$ to $1/4$.

Performing the EOB mapping procedure given in \cite{Damour:2001tu,Barausse:2009xi,Barausse:2011ys}, one will obtain the EOB dynamics that describes a test particle moving in a deformed Kerr-like background. In Appendix \ref{app_A} we summarize the procedure for building the EOB mapping. In the Boyer-Lindquist coordinates $x^{\mu}=(t,r,y,\phi)$, the metric of the deformed Kerr-like spacetime can be formally written as 
\be\label{eff_met}
ds^2_{\rm eff}=
g_{tt}dt^2+g_{rr}dr^2+g_{yy}dy^2+g_{\phi\phi}d\phi^2+2g_{t\phi}dtd\phi\,.
\ee
For the sake of later convenience, we write down the inverse metric $g^{\mu\nu}$, whose components can be expressed as
\bea
&& g^{tt} = -\frac{g_{\phi\phi}}{g_{t \phi}^2 -g_{tt} g_{\phi\phi}}\,,\quad g^{t\phi}=\frac{g_{t\phi}}{g_{t \phi}^2 -g_{tt} g_{\phi\phi}}\,,\quad g^{\phi\phi}=-\frac{g_{tt}}{g_{t \phi}^2 -g_{tt} g_{\phi\phi}}\,,
\nonumber\\
&& g^{rr}=\frac{1}{g_{rr}}\,,\quad g^{yy}=\frac{1}{g_{yy}}\,.\nonumber
\eea
The explicit forms of these components of the inverse metric $g^{\mu\nu}$ will be given later, where they are parameterized by the total mass $M$, symmetric mass ratio $\eta$, and the spin parameter defined as
\be
a:=\frac{|{\bf S}_1 + {\bf S}_2|}{M}
\ee
with ${\bf S}_{1,2}$ the component spins of the binary. Note that $a\le (1-2\eta) M$ in principle due to $|{\bf S}_i|\le m_i^2$, i.e., the fulfillment of the Kerr bound for each binary component. It reduces to the extremality bound for Kerr black holes, $a\le M$ for $\eta=0$.

The binary dynamics is then connected to a probe dynamics of mass $\mu:=m_1 m_2/M$ moving in the above background ``effective metric''. It can be summed up to the following effective Hamiltonian 
\be\label{Heff}
H_{\rm eff}= \beta^i p_i + \alpha \sqrt{\mu^2 +\gamma^{ij} p_i p_j + \cdots} + H^S
\ee
where $\alpha:=1/\sqrt{-g^{tt}}$, $\beta^i:=g^{ti}/g^{tt}$ and $\gamma^{ij}:=g^{ij}-g^{ti}g^{tj}/g^{tt}$ are respectively the lapse function, shift vector, and the reduced 3-metric constructed from the ``effective metric''; the $\cdots$ denotes contributions from higher than 2PN orders. We thus see that the first two terms in \eq{Heff} describe a non-spinning particle of mass $\mu$ moving in the effective background with metric given by \eq{eff_met}. The remaining $H^S$ encodes the spin-orbital interaction and spin-spin self-interaction of the test-body's spin ${\bf S}^*$. In \cite{Barausse:2009xi,Barausse:2011ys} (see also Appendix \ref{app_A}), it has been shown that ${\bf S}^*$ is related to the component spins of the binary, ${\bf S}_{1,2}$ by \cite{Barausse:2009xi,Barausse:2011ys}
\be\label{probe_spin}  
{\bf S}^*:= {m_2\over m_1} {\bf S}_1  + {m_1 \over m_2}{\bf S}_2 + {1\over c^2} \bm{\Delta}_{\sigma^*}\,.
\ee
where $\bm{\Delta}_{\sigma^*}$ is an arbitrary vector function.\footnote{The EOB mapping between ${\bf S}^*$ and $a$ (background spin parameter) and component spins ${\bf S}_{1,2}$ is not unique. Different mappings are related by canonical transformations and redefinitions of the Hamilton-Jacobi function \cite{Barausse:2009xi,Barausse:2011ys}. However, in either EOB frame, we can have ${\bf S}^*:= {m_2\over m_1} {\bf S}_1  + {m_1 \over m_2}{\bf S}_2$ by setting the arbitrary function $\bm{\Delta}_{\sigma^*}$ to some appropriate specific form.}

The key ingredient in \cite{Buonanno:1998gg} obtaining the EOB dynamics is to further map $H_{\rm eff}$ to the so-called EOB-Hamiltonian $H_{\rm EOB}$ for the EOB dynamics. The map is inherited from \eq{E_EOB}, that is,
\be\label{EOB_map}
H_{\rm EOB}=M \sqrt{1+ {2\eta \over \mu}(H_{\rm eff}-\mu)}\,.
\ee
With this identification, the resulting EOB dynamics is natural without the need to introduce additional parameters when performing the EOB mapping. For conservative dynamics without considering the radiation back reaction, we will soon see that the EOB dynamics of $H_{\rm EOB}$ is equivalent to the one of $H_{\rm eff}$.

In this paper, we would like to focus on the effect of the deformed Kerr-like background on the (non-)integrability of the inspiral dynamics. Thus, we will impose the dynamical constraint ${\bf S}^*=0$ by choosing an appropriate $\bm{\Delta}_{\sigma^*}$ so that $H^S=0$. Note that $H^S$ in \eq{Heff} will yield  Mathisson-Papapetrou-Dixon (MPD) equation ${D S^{\mu\nu} \over d\tau} = P^{\mu}u^{\nu}-P^{\mu}u^{\nu}$ with the linear momentum $P^{\mu}=\mu u^{\mu}+{\cal O}(({\bf S}^*)^2)$ with $u^{\mu}$ the 4-velocity of the probe particle. After imposing supplementary spin conditions (e.g. see \cite{Steinhoff:2015ksa} for interesting discussions), the MPD equation can be reduced to ${D {\bf S}^* \over d\tau} = {\cal O}(({\bf S}^*)^2)$ \cite{Faye:2006gx}. Therefore, if we choose ${\bf S}^*=0$ right from the beginning by properly tuning \eq{probe_spin}, we can consistently turn off the probe's spin. For simplicity, in this paper, we will consider only the case with ${\bf S}^*=0$ up to 2PN order. That is, we will consider the following effective Hamiltonian
\be\label{H_eff}
H_{\rm eff}= \beta^i p_i + \alpha \sqrt{\mu^2 +\gamma^{ij} p_i p_j}\,.
\ee
This effective Hamiltonian describes the geodesic motion of a non-spinning probe of mass $\mu$ moving in the deformed Kerr background metric \eq{eff_met}.

We will show that the EOB dynamics based on the above effective Hamiltonian will yield resonant islands/Birkhoff islands on the Poincar\'e surfaces of section even for the probe without spin. The resonant islands/Birkhoff islands we will obtain are due to the deformation from the exact Kerr metric contributed by the non-zero symmetric mass ratio $\eta$. This is in contrast to the chaotic orbits of a spinning probe moving around a Kerr black hole \cite{Kiuchi:2004bv,Han:2008zzf}. The investigation of inspiraling dynamics similar to that of Eq.~\eqref{H_eff} has also been performed \cite{Zhang:2020rxy,Zhang:2021fgy}, in which the authors considered extreme mass-ratio limit, i.e., $\eta\ll1$, and focused on equatorial-eccentric orbits and generic orbits, respectively. In Refs.~\cite{Zhang:2020rxy,Zhang:2021fgy}, the probe spin effects are omitted because of the extreme mass-ratio limit. In addition, in the same limit, one can define an approximated Carter constant, and the system becomes approximately integrable. This is not the case in the scenarios that we will consider here. We will go beyond the extreme mass-ratio limit and investigate how an arbitrary $\eta$ generates chaotic behaviors in the EOB framework.
 
The ``effective metric'' \eq{eff_met} is stationary and axisymmetric, so that $H_{\rm EOB}$ and $p_{\phi}$ are constants of motion, which we will denote as $E_T$ and $L_z$, respectively. In the EOB framework, the reduced mass $\mu$ and the symmetric mass ratio $\eta$ are also dynamically constant. Eq.~\eq{EOB_map} thus implies that $H_{\rm eff}$ is also a constant of motion, which we denote as $E$. Thus, $E_T$ and $E$ are related by
\be\label{ETE0}
E_T=M \sqrt{1+ {2\eta \over \mu}(E-\mu)}\,.
\ee
The Hamilton equations of $H_{\rm EOB}$ are then given by
\bea\label{qdot} 
\frac{dq^i}{dt_{\rm EOB}}&=&{\partial H_{\rm EOB} \over \partial p_i}={M\over E_T} {\partial H_{\rm eff} \over \partial p_i}={M\over E_T}\frac{dq^i}{dt}\,, \\
\label{pdot}
\frac{dp_i}{dt_{\rm EOB}} &=& - {\partial H_{\rm EOB} \over \partial q^i}=-{M\over E_T} {\partial H_{\rm eff} \over \partial q^i}=-{M\over E_T}\frac{dp_i}{dt}\,,
\eea
where we have used \eq{EOB_map} and the constraint
\be\label{EOB_cons}
H_{\rm EOB}=E_T
\ee
to arrive at the second equalities. From the above, i.e., the second equalities in \eq{qdot} and \eq{pdot}, it is clear that we can map the Hamiltonian dynamics of $H_{\rm EOB}$ into the one of $H_{\rm eff}$ by simply rescaling the time coordinate with a factor $E_T/M$, and at the same time replacing the constraint \eq{EOB_cons} by
\be 
H_{\rm eff}=E
\ee
with $E$ relating to $E_T$ by \eq{ETE0}. 

Since the EOB dynamics and effective dynamics are equivalent for finite fixed energy by a simple rescaling of the time coordinate, in the following we will analyze the phase space for the effective dynamics or its associated geodesic dynamics, and the result should also hold for the EOB dynamics. 

\section{(Non-)Integrability of EOB dynamics}\label{sec.pk}
Before solving the aforementioned non-spinning EOB dynamics for the binary Kerr black holes or the equivalent geodesic dynamics in the spacetime described by the effective metric \eq{eff_met}, we would like to examine whether the corresponding system is integrable or not. The fact that  EOB dynamics can be described as the geodesic motion of a test particle around an effective metric allows us to check the integrability of the system by examining the (non)existence of the Carter constant of the effective metric. We follow the criterion given in \cite{1979GReGr..10...79B,Papadopoulos:2018nvd}, which is applied to a metric with two commuting Killing vectors. If a metric satisfies the criterion, the geodesic equations can be separated, and the corresponding dynamics is integrable. The criterion \cite{Papadopoulos:2018nvd} essentially states that the existence of the Carter constant requires the (inverse) metric to take the following Papadopoulos-Kokkotas (PK) form in the Boyer–Lindquist coordinates:  
\bea 
g^{tt}&=&\frac{\mathcal{A}_5(r)+\mathcal{B}_5(y)}{\mathcal{A}_1(r)+\mathcal{B}_1(y)}\,,\qquad g^{t\phi}=\frac{\mathcal{A}_4(r)+\mathcal{B}_4(y)}{\mathcal{A}_1(r)+\mathcal{B}_1(y)}\,, \qquad g^{rr}=\frac{\mathcal{A}_2(r)}{\mathcal{A}_1(r)+\mathcal{B}_1(y)}\,, \nn \\
g^{yy}&=&\frac{\mathcal{B}_2(y)}{\mathcal{A}_1(r)+\mathcal{B}_1(y)}\,, \qquad g^{\phi\phi}=\frac{\mathcal{A}_3(r)+\mathcal{B}_3(y)}{\mathcal{A}_1(r)+\mathcal{B}_1(y)}\,, \label{PK}
\eea 
where $\mathcal{A}_i(r)$ and $\mathcal{B}_i(y)$ are arbitrary functions of $r$ and $y$.

The explicit form of the inverse metric associated with \eq{eff_met} for the 2PN order EOB dynamics can be found in \cite{Barausse:2009xi,Cao:2017ndf}, and we just write down here: 
\bea
g^{tt}&=&-\frac{\Lambda_t}{\Delta_t\Sigma}\,, \qquad g^{t\phi}=-\frac{\tilde\omega_{fd}}{\Delta_t\Sigma}\,,\qquad g^{rr}=\frac{\Delta_r}{\Sigma}\,,\nn\\
g^{yy}&=&\frac{1-y^2}{\Sigma}\,,\qquad g^{\phi\phi}=\frac{1}{\Lambda_t}\left(-\frac{\tilde\omega_{fd}^2}{\Delta_t\Sigma}+\frac{\Sigma}{1-y^2}\right)\,,  \label{EOB_meteric_1}
\eea
where 
\bea
\Delta_t &=& \Delta+\eta F(r)\,,\quad \Delta_r=\Delta_t\left[1+\eta G(r)\right]\,,\quad \tilde\omega_{fd}=a \left(X-\Delta \right) \left[1+\eta H(r)\right]\,, \nn \\
\Lambda_t&=& X^2-a^2\Delta_t\left(1-y^2\right)\,, \label{Delta_t}
\eea
with 
\be  
\Sigma =r^2+a^2y^2\,,\quad \Delta=r^2-2Mr+a^2\,,\quad X=r^2+a^2\,,
\ee 
and 
\bea 
F(r)&=&\frac{2M^3}{r}\,,
%+\frac{M^4}{r^2}ht_1\left(\frac{94}{3}-\frac{41}{32}\pi^2\right)\;, 
\nn \\
 G(r)&=&\frac{1}{\eta} \ln\left[1+6\eta \frac{M^2}{r^2}\right]\,, %+2ht_2\left(26-3\eta\right)\eta \frac{M^3}{r^3}\right]\;, 
 \nn \\
H(r)&=& \frac{1}{2r^2}\left(\omega^{fd}_1 M^2+\omega^{fd}_2 a^2\right)\,. \label{FGH}
\eea 
The parameters $\omega^{fd}_{1,2}$ in the last line of \eq{FGH} are adjustable to regulate the strength of frame dragging. Note that, at 1PN order, the EOB metric \eq{EOB_meteric_1}-\eq{FGH} will simply reduce to the one of the Schwarzschild black hole \cite{Damour:2001tu}. Moreover, one can determine the extremality bound $a_{\rm ext}(\eta)$ by requiring $r \Delta_t(r)=(r-r_1)^2(r-r_2)$ for $r_1>0$ when $\eta\in[0,1/4]$, so that $a_{\rm ext}(\eta)=\sqrt{r_1^2-2r_1 r_2}$. Then, it is straightforward to see that $a_{\rm ext}(\eta) \ge (1-2\eta) M$. Thus, the metric \eq{FGH} always admits an event horizon whenever the Kerr bound holds for both binary components.

Given the above effective metric of EOB dynamics, we can check if it can be recast into the PK form. We immediately can identify 
\bea
\mathcal{A}_1(r)&=& r^2\,,\qquad \mathcal{B}_1(y)=a^2y^2\,,\nn\\
\mathcal{A}_2(r)&=&\Delta_r\,,\qquad \mathcal{B}_2(y)=1-y^2\,,\nn\\
\mathcal{A}_5(r)&=&-\frac{X^2}{\Delta_t}\,,\qquad \mathcal{B}_5(y)=a^2\left(1-y^2\right)\;,\nn\\
\mathcal{A}_4(r)&=&-\frac{a\left(X-\Delta\right)\left[1+\eta H(r)\right]+\Delta_t}{\Delta_t}\,,\qquad \mathcal{B}_4(y)=1\,.
\eea 
This then gives the constraint 
\be
\mathcal{A}_3(r)+\mathcal{B}_3(y)=\frac{1}{1-y^2}-\frac{a^2}{\Delta_t}
-\frac{\eta a^2\Big(X-\Delta_t+(X-\Delta)[1+\eta H(r)]\Big)\Big((X-\Delta) H(r)+ F(r)\Big)}{\Delta_t\Big[X^2-a^2\Delta_t (1-y^2) \Big]}\,.
\label{21constraint}
\ee
The third term involves non-separable $r,y$ dependence so that it is not possible to find appropriate $\mathcal{A}_3(r)$ and $\mathcal{B}_3(y)$ to satisfy the above constraint. However, this term vanishes for either $a=0$ or $\eta=0$. The $\eta=0$ is the probe limit which can be treated as a probe moving around the Kerr black hole whose geodesics are integrable. On the other hand, $a=0$ holds only for the non-spinning binary black holes. The EOB metric becomes a deformed Schwarzschild metric with modified harmonic function but remains spherically symmetric so that the geodesic dynamics is integrable by having an additional Carter constant. Therefore, the generic 2PN EOB dynamics does not contain a second order Killing tensor according to the PK criterion. This is consistent with the results of Ref.~\cite{Zhang:2021fgy}. It should be emphasized that although a second order Killing tensor does not exist, it does not exclude the possibility of the existence of higher order Killing tensors that make the dynamics integrable. Therefore, in order to check non-integrability of the dynamics, we need to resort to numerical integrations of geodesic equations and find chaotic signatures. This is what we will carry out in the next section.

There are general discussions on the integrability of PN inspiral dynamics \cite{Levin:1999zx,Levin:2000md,Hughes:2000nzm,Cornish:2003ig,Hartl:2004xr,Wu:2010mv,2011GReGr..43.2185W,Tanay:2020gfb,Tanay:2021bff}, and it is known that inspiral dynamics up to 1.5PN is integrable. This is consistent with the PK criterion for the EOB dynamics at 1PN, for which the EOB metric is pure Schwarzschild one, i.e., $a=\eta=0$, and at 1.5PN with vanishing test particle spin, for which $a\ne 0$ but $\eta=0$. This suggests that one can deal with the integrability issue of PN inspiral dynamics in the equivalent EOB formalism with the advantage of reducing the two-body phase space to the one-body one.

\section{Chaotic signatures of the EOB dynamics}\label{sec.island}

As discussed in section \ref{EOB_rev} the EOB dynamics based on $H_{\rm EOB}$ of \eq{EOB_map} (up to 2PN order with zero probe spin ${\bf S}^*$) is equivalent to the dynamics based on $H_{\rm eff}$ of \eq{H_eff}. The latter is nothing but the Hamiltonian formulation of a test particle of mass $\mu$ moving in the background geometry described by the metric \eq{EOB_meteric_1}. {Therefore, in this section, we will focus on the geodesic dynamics of a test particle moving in the effective metric \eq{EOB_meteric_1}. Then, we will exhibit the signatures of chaos in the system using the Poincar\'e surfaces of section and the concepts of rotation curves.

\subsection{Theoretical setup: Poincar\'e surface of section and Birkhoff islands}\label{theory}

The dynamics of a massive test particle moving in the effective metric \eq{EOB_meteric_1} can be obtained by solving the geodesic equations of this metric. Because the effective metric is stationary and axisymmetric, namely, the metric functions do not depend explicitly on $t$ and $\phi$, one can define two corresponding constants of motion $E$ and $L_z$. These two constants of motion represent the energy and the azimuthal angular momentum, respectively, and one can use them to obtain the following two equations of motion:
\begin{align}
    \dot{t}&=-g^{tt}E+g^{t\phi}L_z\,,\\
    \dot{\phi}&=-g^{t\phi}E+g^{\phi\phi}L_z\,,
\end{align}
where the dot denotes the derivative with respect to the proper time $\tau$. The evolution of $r$ and $y$ sectors is governed by two coupled second order differential equations, subject to a Hamiltonian constraint, which is associated with the conservation of the rest mass of the test particle. This constraint can be written as
\begin{equation}
    \dot{r}^2+\frac{g^{rr}}{g^{yy}}\dot{y}^2+V_{\rm eff}=0\,,\label{constrainteq}
\end{equation}
where the effective potential reads
\begin{equation}
    V_{\rm eff}=g^{rr}\left(1+g^{tt}E^2+g^{\phi\phi}L_z^2-2g^{t\phi}EL_z\right)\,.
\end{equation}
The values of energy $E$ and azimuthal angular momentum $L_z$ will be normalized with the reduced mass $\mu$ throughout this paper.

Using the constraint equation \eqref{constrainteq}, one can define the so-called the curve of zero velocity (CZV) as
\begin{equation}
    V_{\rm eff}=0\,,
\end{equation}
on which both the radial and angular velocities are zero. CZV determines the boundary of a set of allowed orbits in spacetime. In Figure~\ref{fig:CZV},  we show some typical CZVs in the $(r,y)$ plane for $\eta=0.15$ and different values of spin parameter $a$, $E$ and $L_z$.  For simplicity, we also set $\omega^{fd}_{1,2}$ of \eq{FGH} to zero  in this paper. For this chosen $\eta$, the extremality bound for the metric \eq{FGH} is $a\le a_{\rm ext}(\eta=0.15)=0.847M$, and the requirement from the component Kerr bound gives $a\le (1-2\eta) M=0.7 M$. Figure~\ref{fig:czv1} shows the CZV with $E=0.942$ and $L_z=2.76 M$ in the effective metric with $a=0.67M < 0.7M$, which admits an event horizon at $r=1.60M$.
We see that in this case, the CZV forms a closed region in which the bound orbits reside. Near the event horizon (red dashed line), a second branch of CZV encloses a small region of plunging orbits. Figure~\ref{fig:czv2} shows the CZVs with  $E=0.93671$ and $L_z=1.7542 M$ around two corresponding effective metrics which differ only by the spin $a$, i.e., $a=0.69M$ and $a=0.86M$. The metric for $a=0.69M$ admits an event horizon, but not for $a=0.86M$, i.e., it corresponds to a naked singularity. We see that the CZV around the effective metric with $a=0.86M$ forms a closed curve representing the boundary of a set of bound orbits. On the other hand, the CZV around the effective metric with $a=0.69M$ does not form a closed curve but ends on the horizon. This throat-like CZV connects the region of bound orbits to the plunging region.

\begin{figure}
\centering
\begin{subfigure}{.5\textwidth}
  \centering
  \includegraphics[scale=0.5]{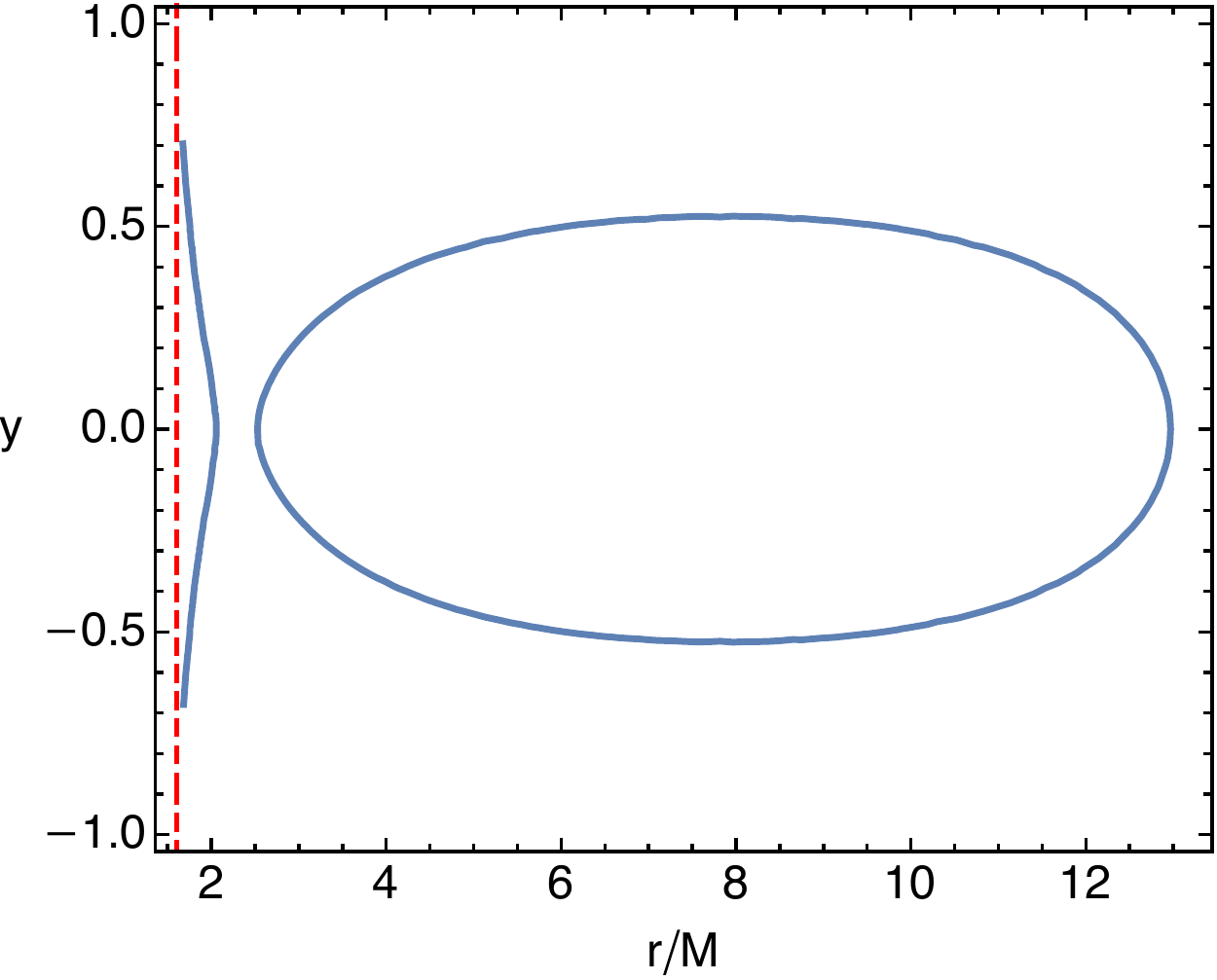}
  \caption{}
  \label{fig:czv1}
\end{subfigure}%
\begin{subfigure}{.5\textwidth}
  \centering
  \includegraphics[scale=0.5]{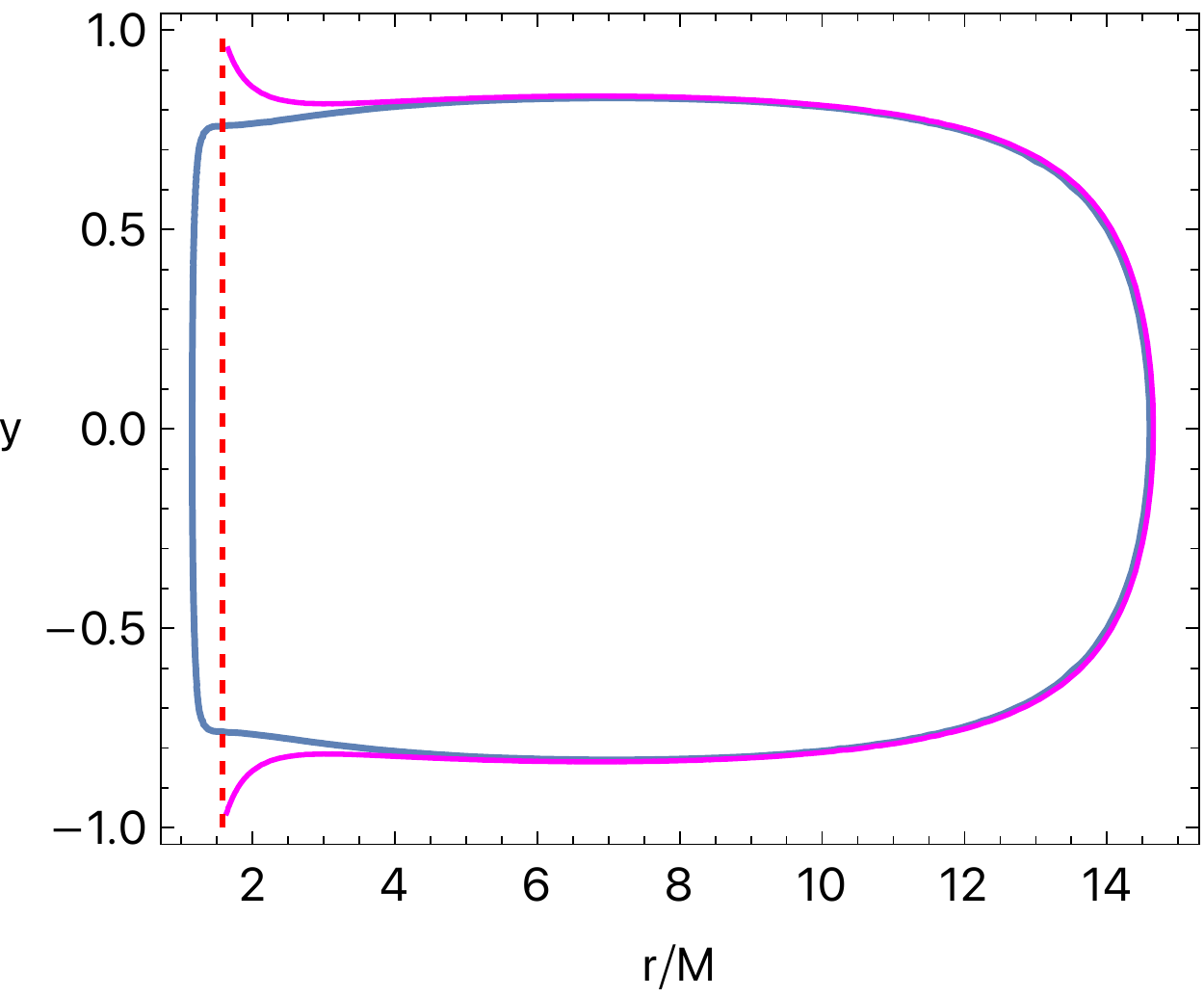}
  \caption{}
  \label{fig:czv2}
\end{subfigure}
\caption{The typical CZVs of the 2PN EOB metrics with $\eta=0.15$ and $\omega^{fd}_1=\omega^{fd}_2=0$ for different values of spin parameter $a$, test particle's energy $E$ and azimuthal angular momentum $L_z$ measured in the unit of reduced mass $\mu$. The event horizon, if it exists, is indicated by the red dashed line. (a) The CZV with $E=0.942$, $L_z=2.76 M$ around the effective metric with $a=0.67M$ forms a closed region, but appears a second non-closed branch near the horizon. (b) Two CZVs with $E=0.93671$ and $L_z=1.7542 M$ around the effective metric spin $a=0.86M$ (blue) and $a=0.69M$ (magenta). Note that the two CZVs here overlap each other for larger $r$ but differ significantly at small $r$. Due to the 2PN correction, the extremality bound for the absence of naked singularity for the EOB metric is no longer $a\le M$ but $a\le a_{\rm ext}(\eta)<M$. For $\eta=0.15$, $a_{\rm ext}=0.847M$. Thus, the metric with $a=0.86M$ does not admit an event horizon, such that the naked singularity appears but the CZV is closed in this case. The CZV with $a=0.69M$ in (b) has a throat-like shape and connects to the horizon. }
\label{fig:CZV} 
\end{figure}

The geodesic equations of test particles moving around a Kerr spacetime are integrable. The integrability of this system results from the existence of a hidden symmetry of the spacetime that allows one to define a Carter constant $\mathcal{K}$. Technically, this constant can be identified as a decoupling constant when separating the radial sector of the geodesic equations from the sector of polar angle. As an integrable system, the phase orbits of particles moving around the Kerr spacetime lie on 2-dimensional tori in phase space. Each torus is characterized by specific values of constants of motion ($E$, $L_z$, $\mathcal{K}$, and the mass of the test particle) and consequently by orbital frequencies and their ratio. In particular, the orbits, in this case, oscillate in both the radial and latitudinal directions, with characteristic frequencies given by $\omega_r$ and $\omega_\theta$, respectively. The ratio of these frequencies is called rotation number $\nu_\theta\equiv \omega_r/\omega_{\theta}$. If the rotation number of an orbit is rational, the orbit is periodic, forming closed curves on the torus. Orbits of this type are called resonant orbits. On the other hand, if $\nu_\theta$ is irrational, the orbits can cover the torus densely and are called quasi-periodic. It turns out that the rotation number can be used to classify orbits. Most importantly, it can also identify the signatures of chaos, even if the chaotic behaviors are very weak.

The integrable geodesic motion bears geometric features in the reduced phase space by constructing the Poincar\'e map, see e.g. \cite{2002ocda.book.....C}. To proceed, one cuts a surface through the foliage of the tori on which the corresponding quasi-periodic orbits densely cover as time goes on, and then each of such tori makes a closed curve on the surface. This surface and the closed curves constructed above are called Poincar\'e surfaces of section and invariant curves, respectively. Typical invariant curves for Kerr geodesics of different values of $\mathcal{K}$ but fixed $E$, $L_z$ and $a$ are shown in Figure~\ref{fig:Poincare_section_Kerr}. The Poincar\'e surface of section shown here is chosen to be the equatorial plane, i.e., $y=0$. We plot the intersections of orbits when they pierce through the equatorial plane with a positive $y$ direction. The black contour is defined by $\dot{y}=0$ and shows the boundary of bound orbits. As shown, the invariant curves are continuous and different curves corresponding to different values of $\mathcal{K}$ are nested within each other \cite{Verhaaren:2009md}.

\begin{figure}[!ht]
  \centering
 \includegraphics[scale=0.5]{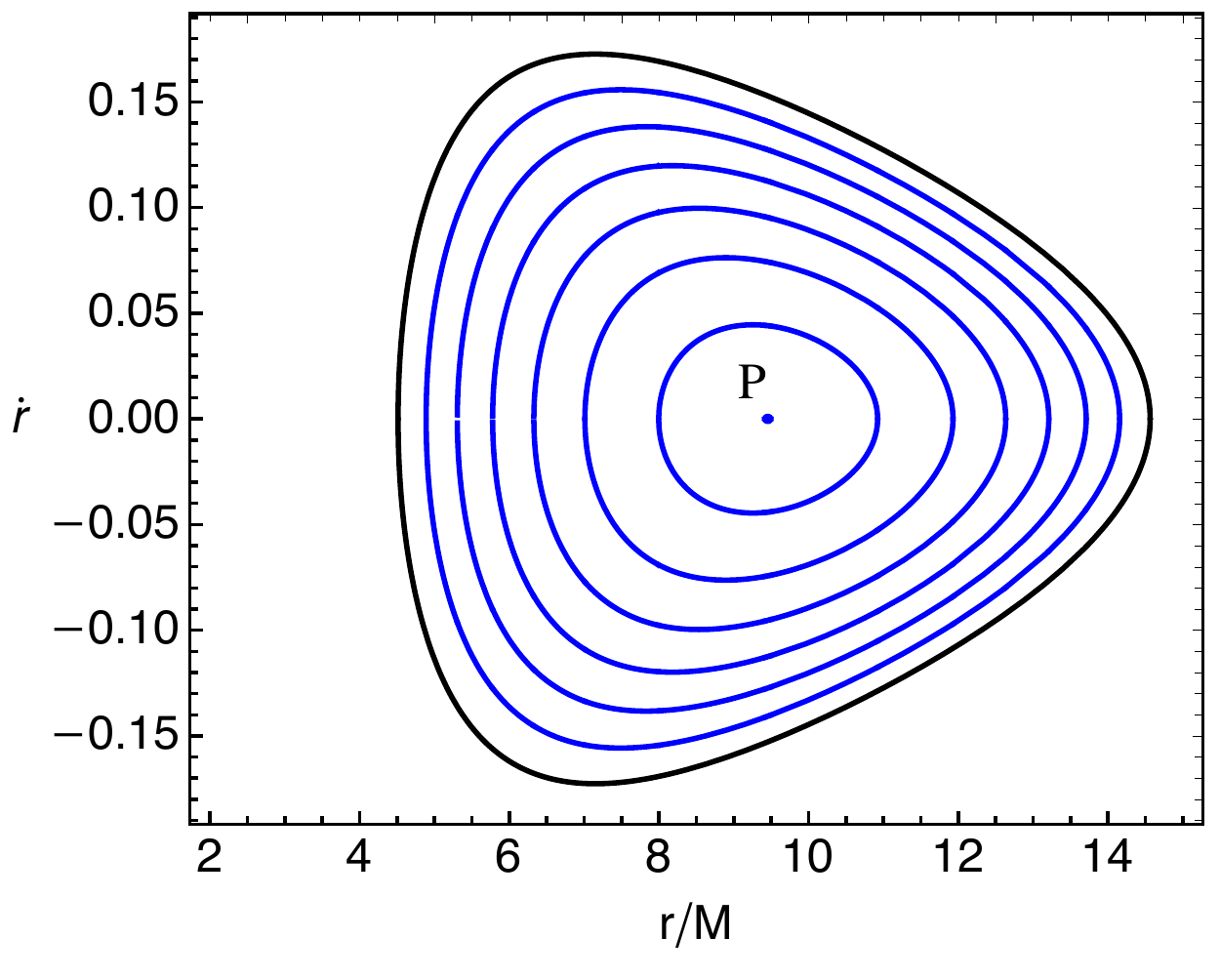}
\caption{Poincar\'e surface of section of the integrable geodesic motion in the Kerr spacetime. Each invariant curve (blue) corresponds to a specific value of the Carter constant. Here, we choose $E=0.95$, $L_z=3M$, and $a=0.9M$.}
\label{fig:Poincare_section_Kerr} 
\end{figure}

If the integrable system is perturbed by non-integrable deformations, as in the case of EOB dynamics which can be treated as geodesic dynamics in a deformed Kerr metric, the system becomes chaotic. Essentially, the behavior of orbits in the phase space depends on whether the deformations are large or not. If the deformations are large, strong chaos may appear, and the whole phase space portrait could be significantly destroyed. However, if the deformations are small, according to the KAM theorem \cite{2002ocda.book.....C,schuster_just_2005}, most of the tori remain undestroyed except for those corresponding to the resonant orbits. Correspondingly, the original invariant curves on the Poincar\'e surfaces of section become slightly deformed but remain continuous. They are called KAM curves.

For resonant orbits, the situations become more complicated. According to the Poincar\'e-Birkhoff theorem \cite{lichtenberg_lieberman_1992,2002ocda.book.....C}, the tori on which resonant orbits are located may dissolve after non-integrable deformations enter. Correspondingly, the resonant points on the Poincar\'e surfaces of section would be split into a number of periodic points, with half of them stable and the other half unstable. Around the stable periodic points, small islands of nested KAM curves appear, which are called Birkhoff chains of islands. For the orbits within the islands, the rotation numbers should be equal to those of the resonant orbits in the original integrable system. The rotation number of an orbit starting with a given initial condition can be operationally obtained by the limiting formula \cite{Voglis_1998,2002ocda.book.....C}:
\begin{equation}\label{rot_num_op}
\nu_{\theta}=\lim_{N\rightarrow \infty} \frac{1}{2\pi N}\sum_{i=1}^{N}\vartheta_i\,,
\end{equation}
where $\vartheta_i$ is the angle between the position vectors of two successive piercings of the given orbit on the Poincar\'e surface of section. The position vectors are defined with respect to the central fixed point $P$, which is defined by the unique piercing of the orbit that only crosses through the surface once (We show a central point $P$ in Figure~\ref{fig:Poincare_section_Kerr}). In practice, we can obtain a well-approximated value of $\nu_{\theta}$ for large enough $N$.
The appearance of Birkhoff islands in Poincar\'e surfaces of section is a clear signature of chaos in the system, and it will result in rational plateaus in the rotation curve. That is, $\nu_{\theta}$ will show plateaus at rational values such as $1/2$, $2/3$, etc, when varying the initial conditions.

\subsection{Birkhoff islands of EOB orbits}\label{2PN_chaos}
In this section, we investigate the breaking up of tori in a Poincar\'e section of the phase space, treating the geodesic motion in the effective metric of \eq{EOB_meteric_1} as a perturbation to the integrable Kerr system. As mentioned in section \ref{EOB_rev}, the effective metric \eq{EOB_meteric_1} is parameterized by mass $M$, spin parameter $a$, and symmetric mass ratio $\eta$. Therefore, different values of these parameters along with the choice for energy $E$ and angular momentum $L_z$ of the test particle may affect the degree of chaos exhibited by trajectories of phase orbits on Poincar\'e surfaces of section. To identify chaotic features using this approach, we numerically integrate the Hamilton equations given by the geodesic dynamics of the effective metric \eqref{EOB_meteric_1}. While solving the equations, we keep the constraint equation \eqref{constrainteq} being satisfied down to $10^{-8}$ during the whole integration using the standard Runge-Kutta eighth order method. Each trajectory is integrated till its characteristic patterns on the Poincar\'e surface of section robustly appear, e.g. continuous KAM curves, clear Birkhoff chains of islands, etc. The integration time varies among different trajectories. Normally, the integration lasts for up to $10^3\sim 10^4$ periods for each trajectory.

\begin{figure}
\centering
\begin{subfigure}{.5\textwidth}
  \centering
  \includegraphics[width=0.85\linewidth]{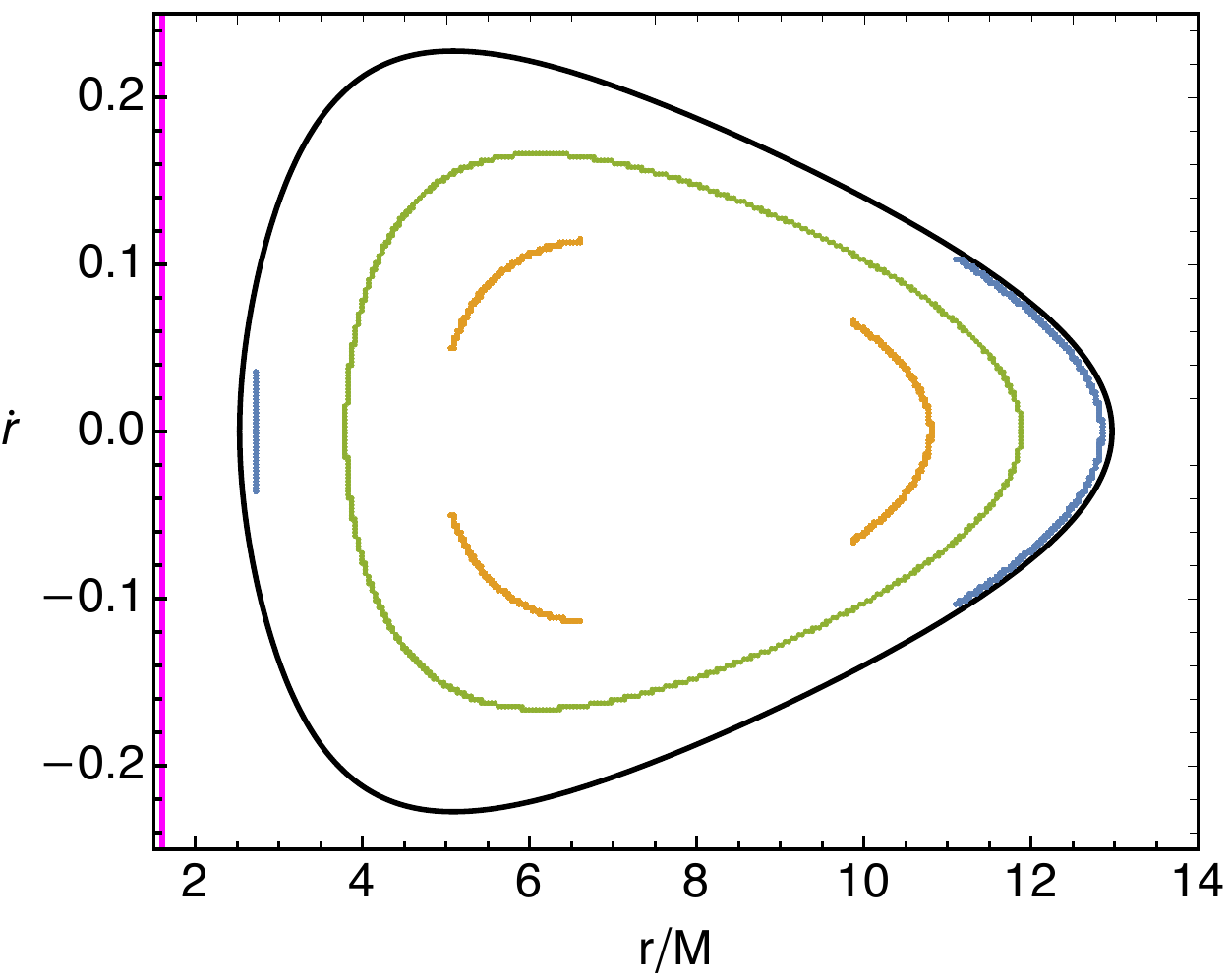}
  \caption{}
  \label{fig:horizonisland}
\end{subfigure}%
\begin{subfigure}{.5\textwidth}
  \centering
  \includegraphics[width=1\linewidth]{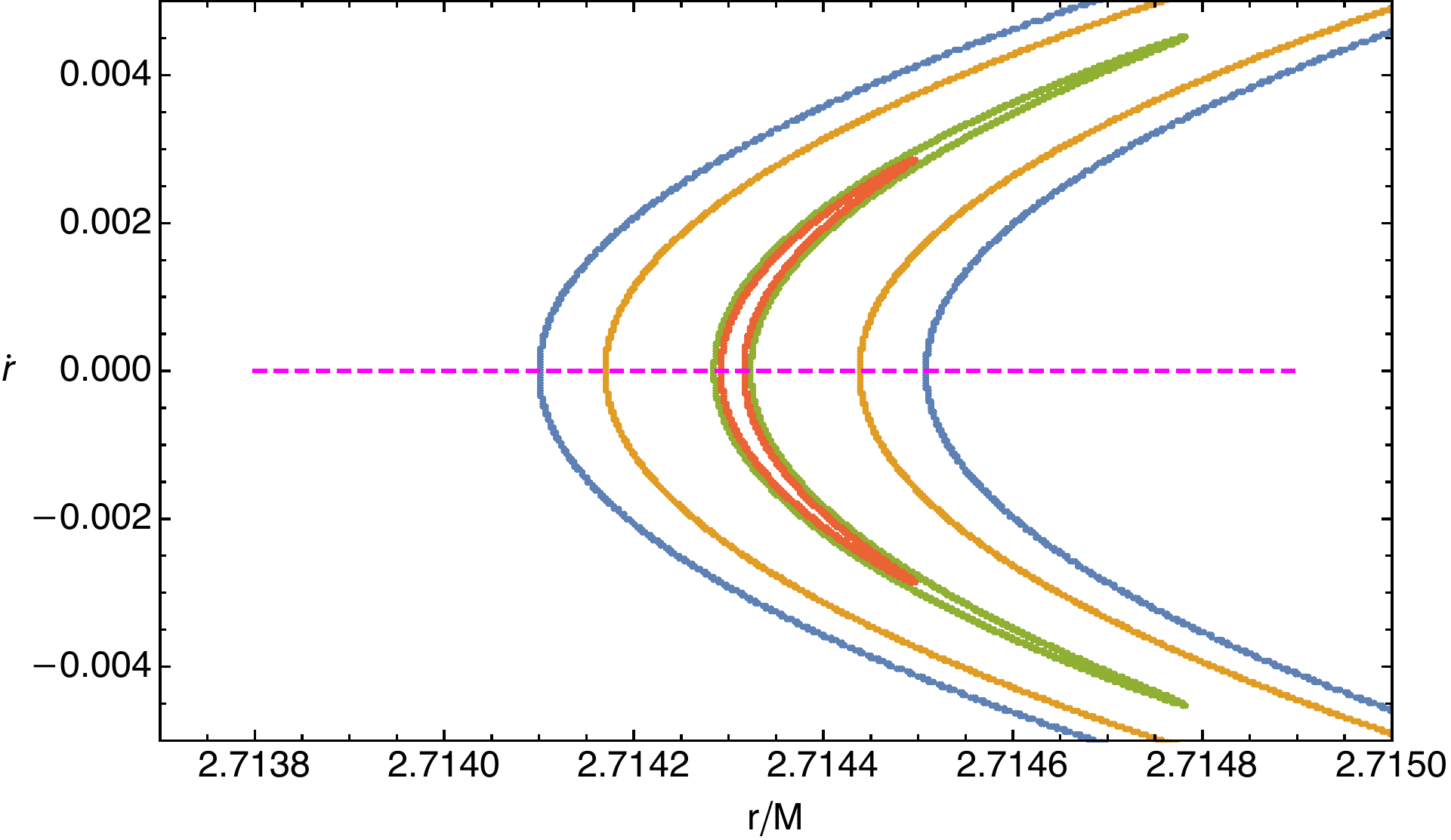}
  \caption{}
  \label{fig:horizonislandzoom}
\end{subfigure}
\caption{Poincar\'e surface of section for geodesic motions in the EOB metric \eq{EOB_meteric_1} for the spin parameter $a=0.67M$, symmetric mass ratio $\eta=0.15$, energy $E=0.942$, and azimuthal angular momentum $L_z=2.76M$. (a) Birkhoff islands (blue: $1/2$-resonance, orange: $2/3$-resonance) and one KAM curve (green). The vertical magenta line indicates the event horizon. (b) Zoom-in on the left branch of the $1/2$-resonant islands.}
\label{fig:islandshorizon}
\end{figure}

We first consider the case where $a<(1-2\eta)M$ such that each of the two binaries satisfies the Kerr bound. More explicitly, we consider the 2PN EOB metric with $a=0.67M$ and $\eta=0.15$, then we set $E=0.942$ and $L_z=2.76 M$. A slight change of $E$ and $L_z$ will not affect the qualitative behaviors of our results. The results are shown in Figure~\ref{fig:islandshorizon}. In Figure~\ref{fig:horizonisland}, one sees a closed KAM curve (green), which is associated with an irrational-$\nu_\theta$ quasi-periodic orbit. In addition, two sets of chains of Birkhoff islands can be clearly identified. They correspond to rational-$\nu_\theta$ resonant orbits. In Figure~\ref{fig:horizonisland}, we show the islands that correspond to $1/2$ (blue) and $2/3$-resonant orbits (orange). The detailed structure of the left branch of the $1/2$-resonant islands is exhibited in the zoom-in Figure~\ref{fig:horizonislandzoom}. The Birkhoff islands come from the splitting resonant points with rational $\nu_{\theta}$, and chaotic behaviors should be characterized by the discontinuity of $\nu_{\theta}$ when varying the initial conditions, as suggested by the KAM theorem and Poincar\'e-Birkhoff theorem.

Using the operational method of \eq{rot_num_op}, we can specifically consider a collection of orbits whose initial conditions are parameterized by $(r_0,y_0,\dot r_0)=(r,0,0)$. The initial points of these orbits correspond to a set of piercings of the orbits through the equatorial plane with zero radial velocities. This set of piercings appears as a horizontal line on the Poincar\'e surface of section. We can then draw the so-called rotation curve, which is defined by the rotation numbers as a function along with this set of initial conditions, that is, as a function of $r$. In Figure~\ref{fig:rotnbhorizon1}, we choose a set of orbits whose initial points correspond to the magenta dashed line in Figure~\ref{fig:horizonislandzoom} and calculate their rotation numbers. As one can see from this figure, when the orbits are inside the island, their rotation numbers are nearly a constant, forming a plateau in the rotation curve. More explicitly, we see that the rotation number monotonically increases with $r$ except for a finite range of radii in which $\nu_\theta$ remains unchanged. The constant value of rotation number, i.e., $\nu_{\theta}=1/2$ at the plateau, indicates that the islands in blue in Figure \ref{fig:horizonisland} correspond to the `$1/2$' resonance of phase orbits. On the other hand, in Figure~\ref{fig:rotnbhorizon2}, we consider a set of orbits whose initial points form a horizontal line on Figure~\ref{fig:horizonisland} that crosses the upper-left branch of the $2/3$-resonant islands. The rotation curve has a clear plateau on which the rotation number remains at $\nu_\theta=2/3$. The existence of Birkhoff islands and plateaus in the rotation curve are clear signatures of chaos.

\begin{figure}
\centering
\begin{subfigure}{.5\textwidth}
  \centering
  \includegraphics[width=0.95\linewidth]{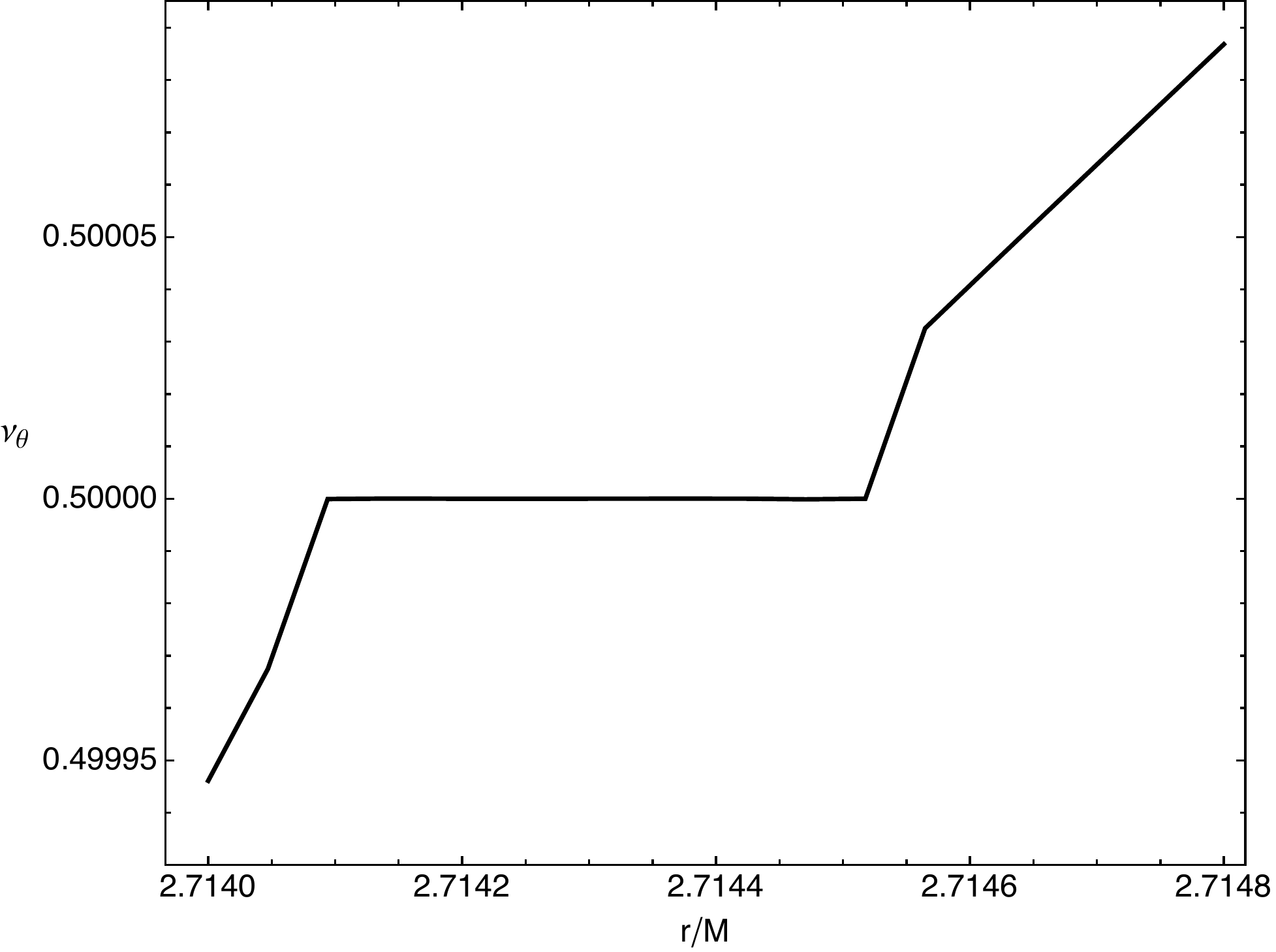}
  \caption{}
  \label{fig:rotnbhorizon1}
\end{subfigure}%
\begin{subfigure}{.5\textwidth}
  \centering
  \includegraphics[width=0.95\linewidth]{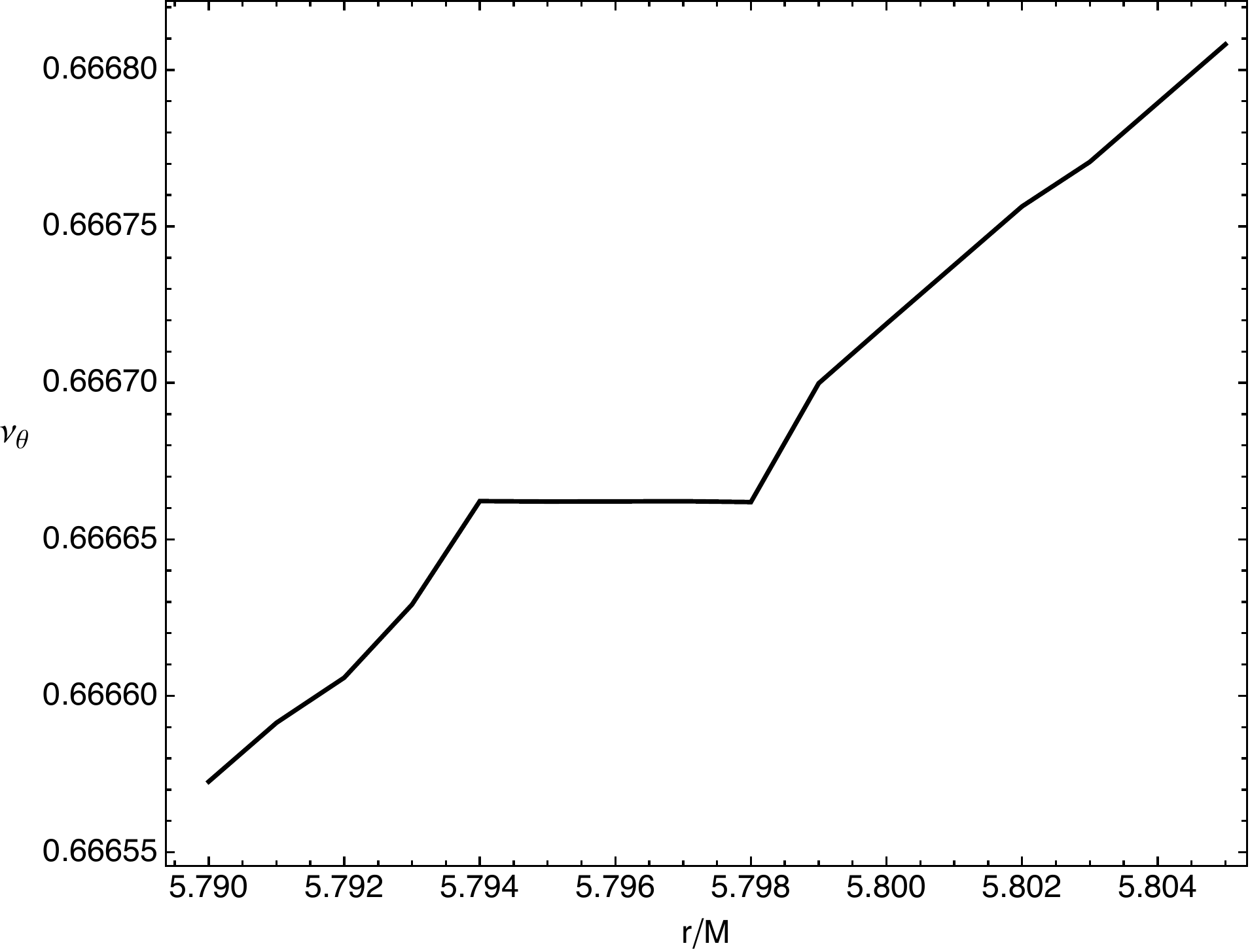}
  \caption{}
  \label{fig:rotnbhorizon2}
\end{subfigure}
\caption{(a) The rotation curve drawn along the magenta dashed line in Figure~\ref{fig:horizonislandzoom}. The plateau has a constant rotation number $\nu_\theta=1/2$ and corresponds to the $1/2$-resonant Birkhoff islands shown in Figure~\ref{fig:islandshorizon}. (b) The rotation curve drawn along a horizontal line in Figure~\ref{fig:horizonisland} that crosses the upper-left branch of the $2/3$-resonant islands. The plateau appears as shown in both cases because the rotation number remains a constant when crossing an island.}
\label{fig:rotationnbhorizon}
\end{figure}

While scanning the parameter space, we find that with higher values of spin parameter $a$ and symmetric mass ratio $\eta$, the chaotic tendency of phase orbits becomes stronger. The effects would be even more profound when the extremality bound is violated, i.e., $a>a_{\rm ext}(\eta)$. This motivates us to investigate in more details the phase orbits in these configurations. Although these configurations may not be strictly physical due to the existence of naked singularities, they can provide us with some insights about how the extremely rapid spins could enhance the chaotic nature of the system. Therefore, we now consider the 2PN EOB metric with $a=0.92M$ and $\eta=0.15$. We hereafter will set the fixed values for $E=0.9367$ and $L_z=1.7542M$ for the 2PN cases. Again, a slight change of these constants of motion does not affect the qualitative results discussed below. The results are shown in Figure~\ref{fig:islands2}.

In Figure \ref{fig:islands2}, we can see the features of KAM curves associated to the irrational-$\nu_{\theta}$ quasi-periodic orbits and obtained by different initial conditions, i.e., the closed curves with different colors (orange/green). In addition, in the zoom-in Figure \ref{fig:sub22} we can again clearly see the features of Birkhoff islands associated with rational-$\nu_{\theta}$ resonant orbits. As the strength of the non-integrability is characterized by the combined effect of $a$ and $\eta$, we later will see that, with fixed $E$ and $L_z$, the strength of chaotic behaviors of the system would be enhanced when $a$ and $\eta$ increase.

\begin{figure}
\centering
\begin{subfigure}{.5\textwidth}
  \centering
  \includegraphics[width=0.85\linewidth]{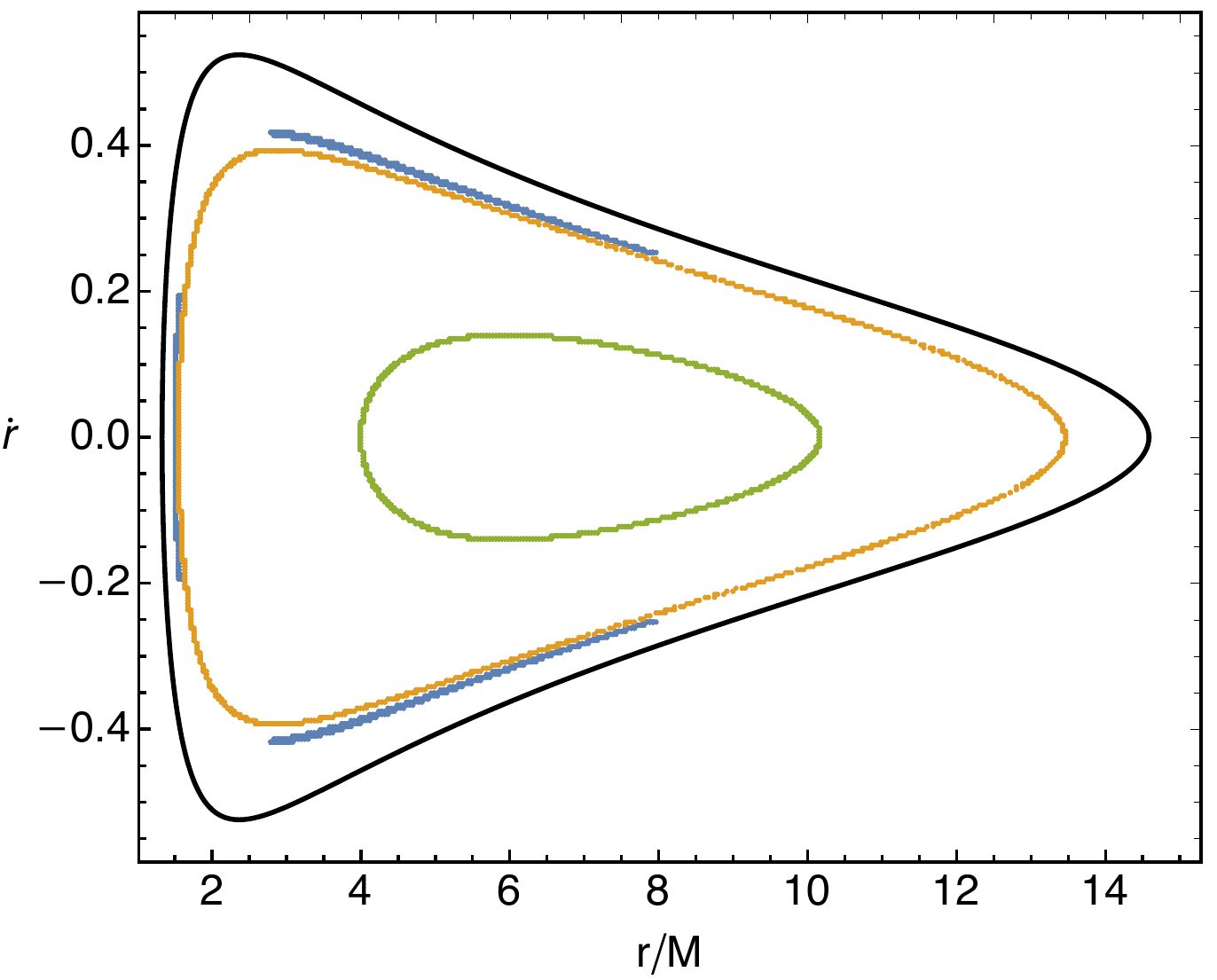}
  \caption{}
  \label{fig:sub21}
\end{subfigure}%
\begin{subfigure}{.5\textwidth}
  \centering
  \includegraphics[width=0.85\linewidth]{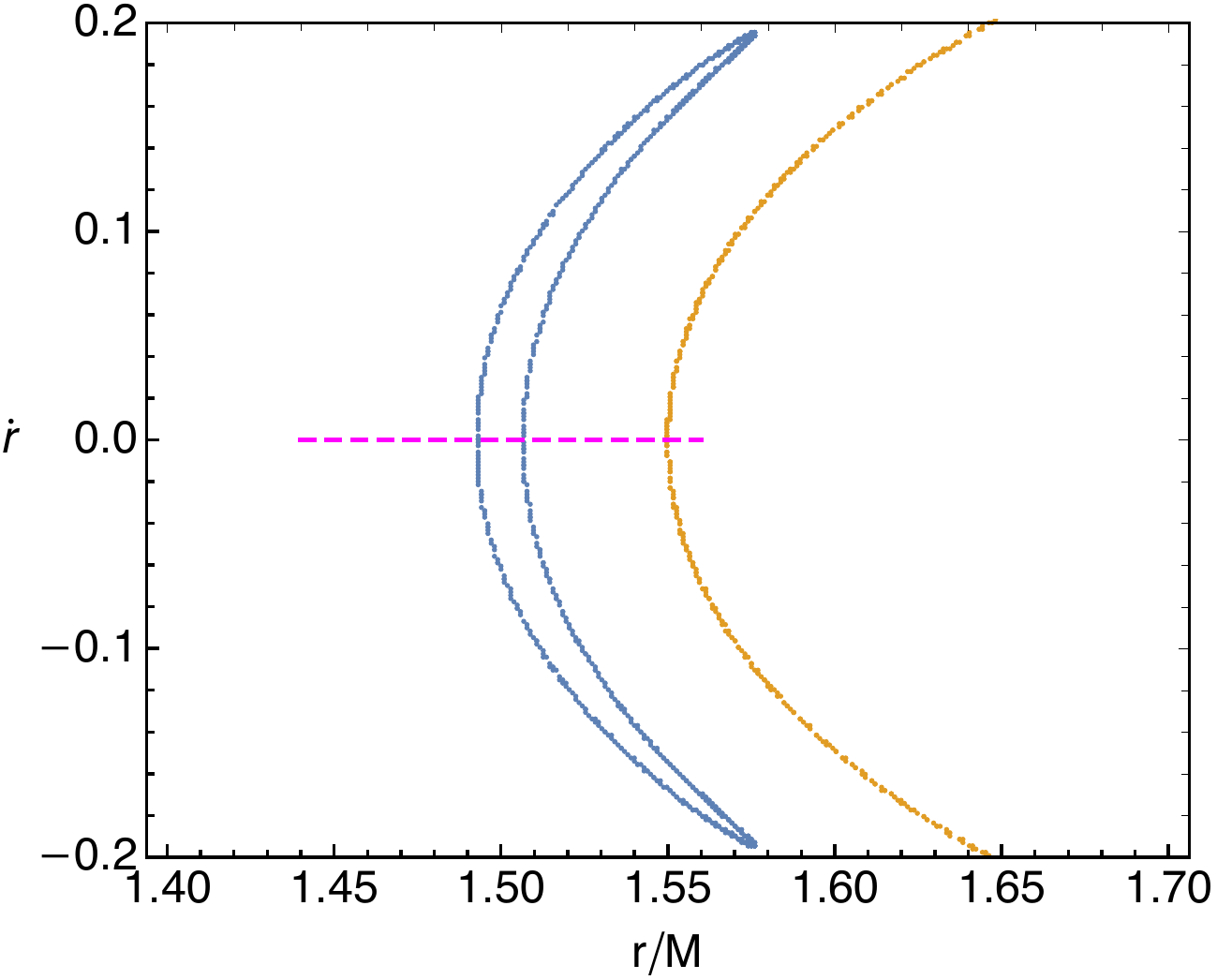}
  \caption{}
  \label{fig:sub22}
\end{subfigure}
\caption{Poincar\'e surface of section for geodesic motions in the EOB metric \eq{EOB_meteric_1} for the spin parameter $a=0.92M$, symmetric mass ratio $\eta=0.15$, energy $E=0.9367$, and azimuthal angular momentum $L_z=1.7542M$. (a) KAM curves and Birkhoff islands ($2/3$-resonance) for initial conditions $(r_0,y_0,\dot{r}_0)=(1.493M,0.1,0)$ (blue), $(1.55M,0,0)$ (orange), and $(4M,0,0)$ (green). (b) Zoom-in on the $2/3$-resonant Birkhoff island (blue) with initial condition $(r_0,y_0,\dot{r}_0)=(1.493M,0.1,0)$. A nearby KAM curve is also shown (orange).}
\label{fig:islands2}
\end{figure}

In Figure \ref{fig:rotnum2}, we choose a set of orbits whose initial conditions correspond to the magenta dashed line in Figure \ref{fig:sub22}, and calculate their rotation numbers. As one can see from this figure, when the orbits are inside the island, i.e., the region enclosed by the blue curve in Figure \ref{fig:sub22}, the rotation curve has a plateau. Indeed, the rotation number monotonically decreases with $r$ except for a finite range of radius around $r=1.5 M$ in which $\nu_\theta$ remains unchanged. The constant value of rotation number i.e. $\nu_{\theta}=2/3$ at the plateau indicates that the three branches of islands (shown in blue) in Figure \ref{fig:islands2} correspond to the `$2/3$' resonance of phase orbits. 

\begin{figure}[!ht]
  \centering
 \includegraphics[scale=0.5]{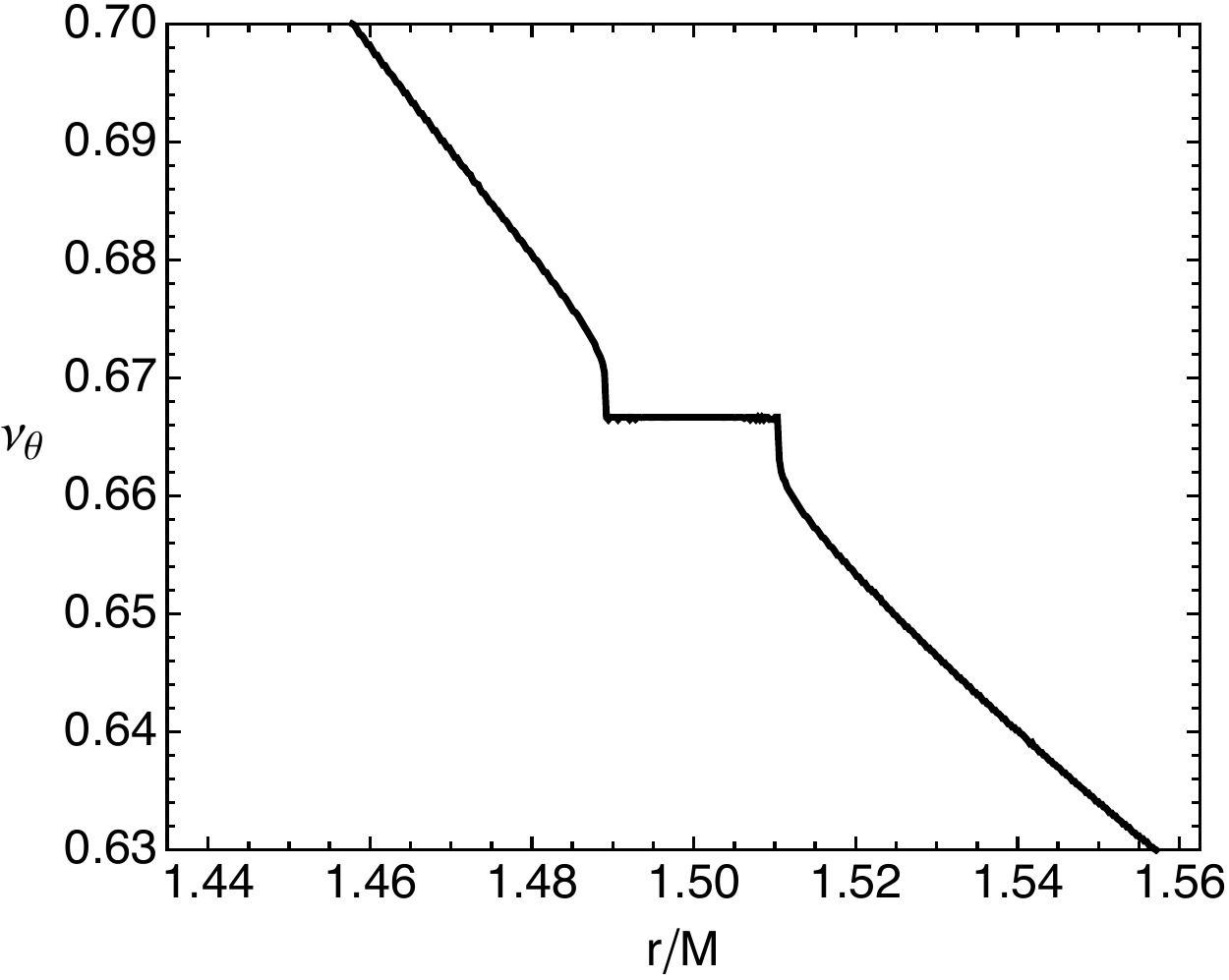}
\caption{The rotation curve drawn along the magenta dashed line in Figure \ref{fig:sub22}, in which $\eta=0.15$ and $a=0.92M$. The plateau around $r=1.50 M$ corresponds to the $2/3$-resonant Birkhoff island shown in Figure \ref{fig:sub22}.}
\label{fig:rotnum2} 
\end{figure}

In addition to plateaus that indicate the existence of Birkhoff islands, there is another feature in rotation curves that also indicates chaos in a dynamical system. In Figure \ref{fig:rotnum22}, we consider the same EOB metric and constants of motion $E$  and $L_z$ as those in Figure \ref{fig:rotnum2}, but draw the rotation curve in a different range of $r$. As one can see in this figure, instead of a plateau, the rotation curve could be featured by an abrupt change. In this case, the abrupt change at $\nu_\theta=2/3$ corresponds to the region near an unstable point. Near this region, the tori structure is only deformed, compared with the original invariant curves, but not yet destroyed.   

\begin{figure}[!ht]
  \centering
 \includegraphics[scale=0.5]{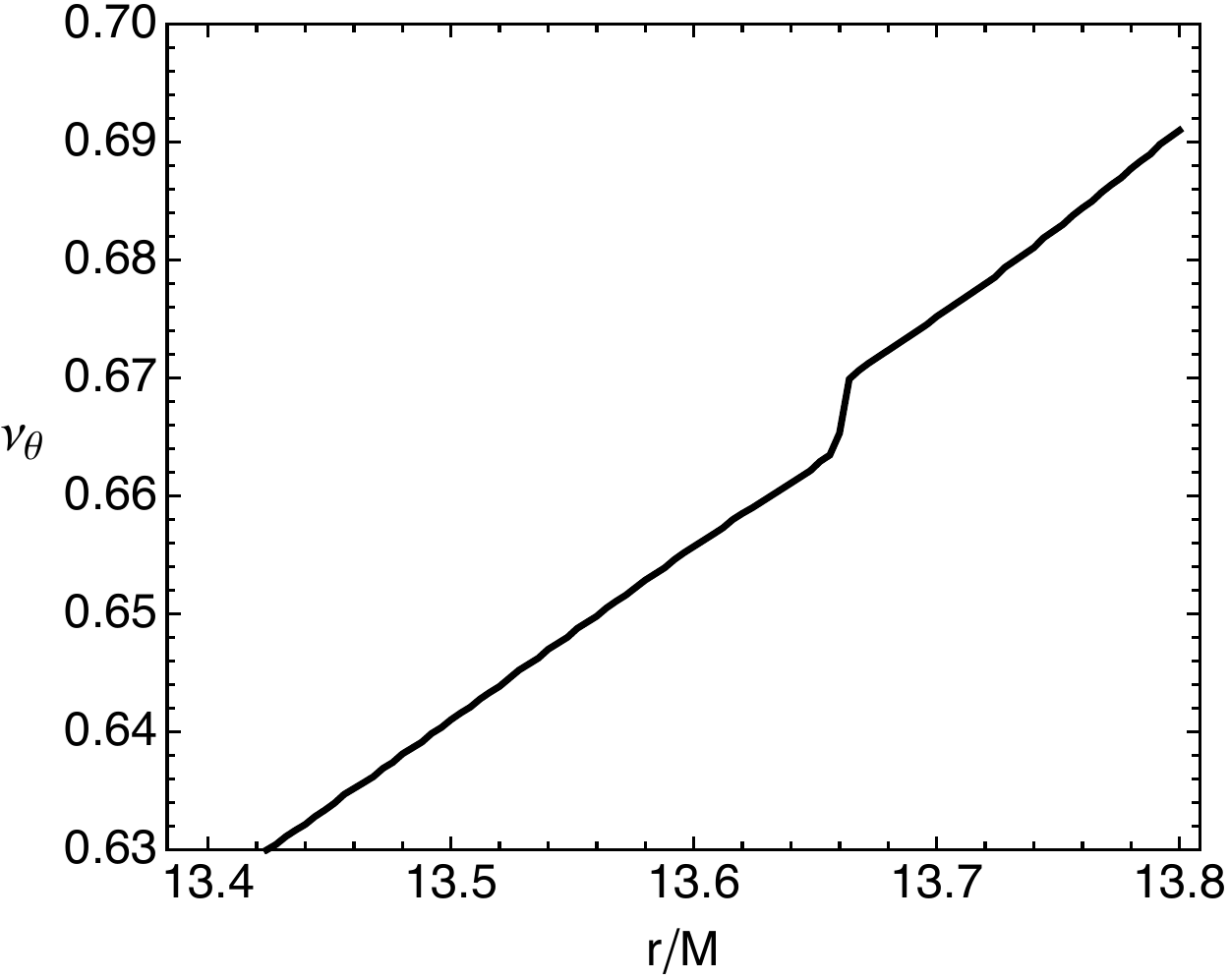}
\caption{The abrupt change in the rotation curve at $\nu_\theta=2/3$ corresponds to the region near an unstable periodic orbit.}
\label{fig:rotnum22} 
\end{figure}

\begin{figure}[!ht]
  \centering
 \includegraphics[scale=0.6]{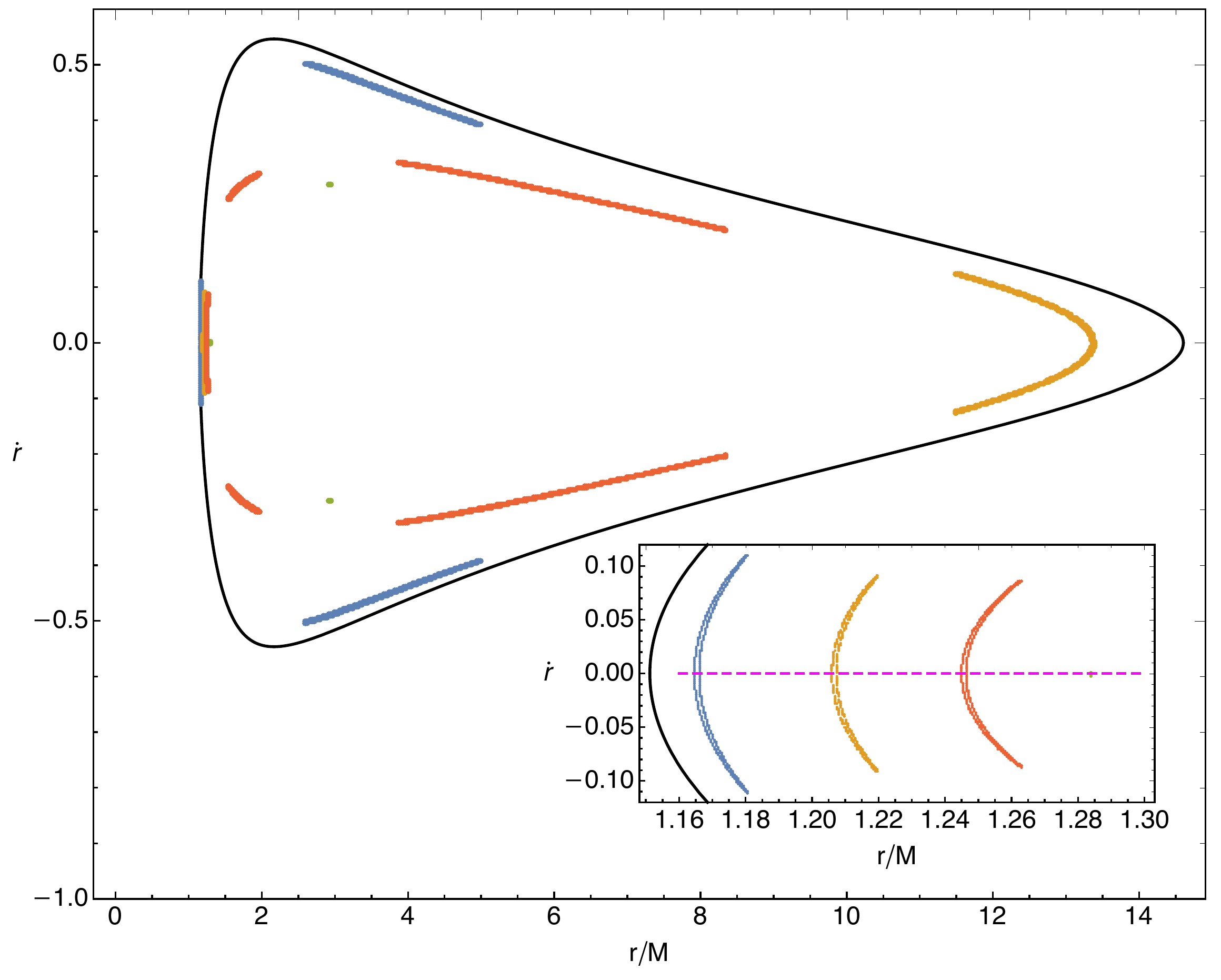}
\caption{Birkhoff chains of islands for 2PN Hamiltonian for $\eta=0.15$ and $a=0.86M$. The inset shows the zoom-in of the islands on the left edge of the Poincar\'e surface of section, which consists of $2/3$ (blue), $1/2$ (orange), $2/5$ (red), and $1/3$-resonances (green).}
\label{fig:Birkhoff_island_2PN_0.15eta_0.86spin} 
\end{figure}

In fact, rotation curves can be used not only as a tool to search for chaos in the dynamical system, but also to locate the Birkhoff islands in the Poincar\'e surfaces of section. In particular, the islands that correspond to higher-order resonances are typically very thin and are difficult to identify. For example, one can first calculate the rotation curve along the whole horizontal axis ($\dot{r}=0$) within CZV, and find the short range of radius $r$ that is associated to a rational $\nu_\theta$. These radii should correspond to the initial conditions $r_0$ of the orbits that cross the islands if islands exist. This method can substantially assist in finding islands of high resonances. In Figure \ref{fig:Birkhoff_island_2PN_0.15eta_0.86spin} we consider $a=0.86M$ and $\eta=0.15$, then exhibit the Birkhoff islands of $2/3$ (blue), $1/2$ (orange), $2/5$ (red), and $1/3$-resonances (green). The inset shows the zoom-in of these islands near the left edge of the Poincar\'e surface of section. The rotation curve along the horizontal magenta line is shown in Figure \ref{fig:rotationnumber_2PN_0.15eta_0.86spin}. The plateaus appear when the orbits cross the islands, with the rotation numbers being $2/3$, $1/2$, $2/5$, and $1/3$, from left to right, respectively.

\begin{figure}
\centering
  \includegraphics[width=0.65\linewidth]{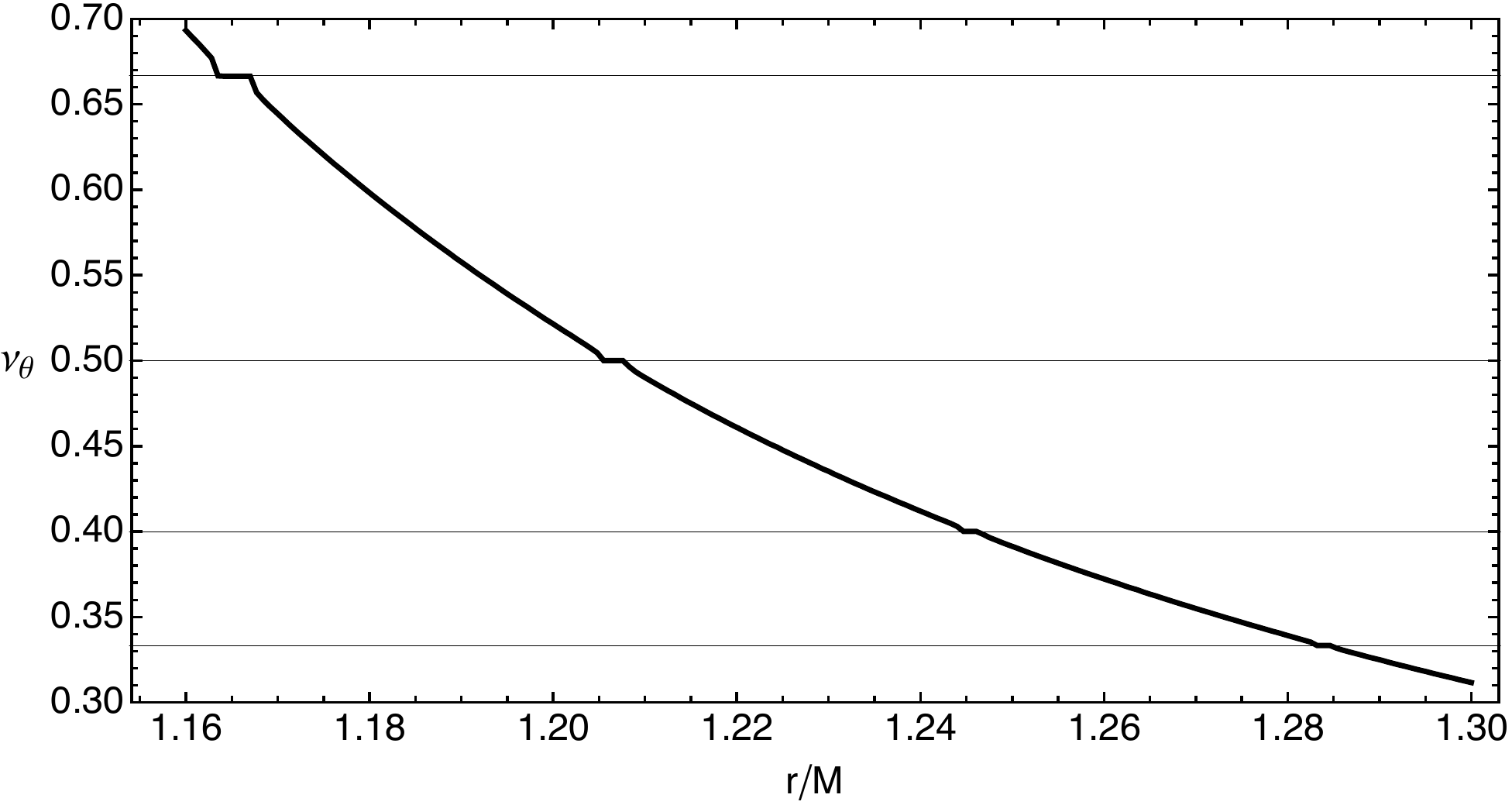}
\caption{The rotation curve drawn along the horizontal magenta line in the inset of Figure \ref{fig:Birkhoff_island_2PN_0.15eta_0.86spin}. Plateaus that correspond to different orders of resonances are clearly visible.}
\label{fig:rotationnumber_2PN_0.15eta_0.86spin}
\end{figure}

\begin{figure}[!ht]
  \centering
 \includegraphics[scale=0.6]{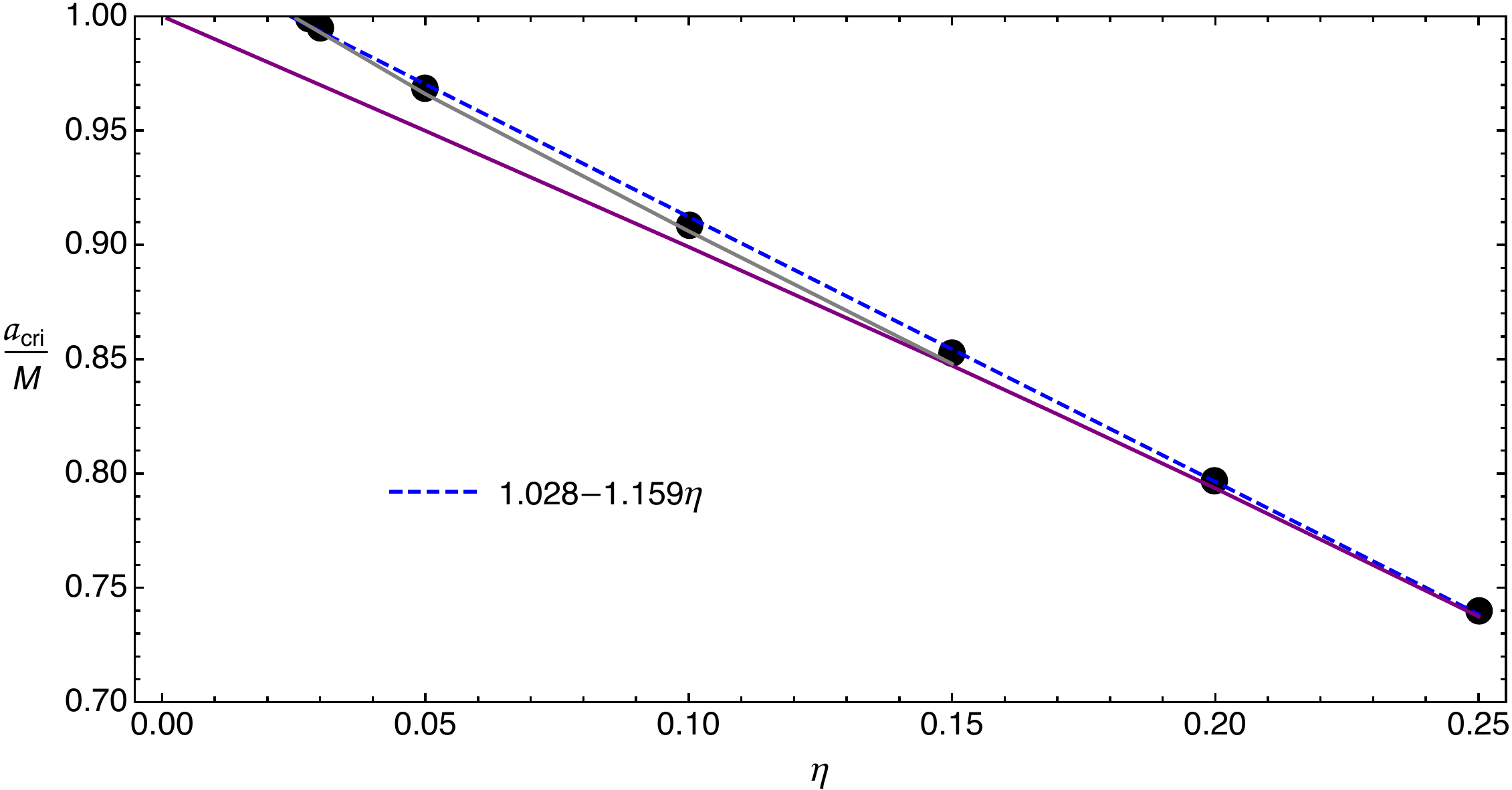}
\caption{The critical spin $a_{\rm cri}$ for some values of the symmetric mass ratio $\eta$ (black points) upon fixing $E=0.9367$ and $L_z=1.7542M$. The points fit very well with a linear function (blue dashed line). Below the blue dashed line or the gray curve, the leftmost $2/3$-resonant island either disappears or becomes too small to measure. Also, the extremality bound $a_{\rm ext}(\eta)$ is shown by the purple curve. Note that $a_{\rm cri}$ as a function of $\eta$ depends also on the choice of $E$ and $L_z$. We can also observe that the case with closed CZV in Figure \ref{fig:czv2} lies above the blue dashed line and one with open CZV resides below it. For this particular choice of $E$ and $L_z$, the above three curves are near to each other which makes it interesting to explore the role played by the event horizon in formation or destruction of islands.}
\label{fig:criticalspin} 
\end{figure}

Before closing this subsection, we would like to investigate how the strength of chaos is affected by the interplay of the metric parameters $a$ and $\eta$ when fixing $E$ and $L_z$. This can be done using the Poincar\'e surface of section. Firstly, we fix $E=0.9367$ and $L_z=1.7542M$, then we consider the $\{a,\eta\}$ configuration where the extremality bound is violated. We choose this configuration because its island structures are more apparent and easier to identify. We focus on the leftmost branch of the $2/3$-resonant islands, e.g. the blue islands in Figure \ref{fig:islands2}. This branch of islands is always vertically symmetric with respect to the $\dot{r}=0$ axis. Therefore, it is convenient to use the width on the $\dot{r}=0$ axis of this branch of islands as a measure of the strength of chaos in the system. For a given value of the symmetric mass ratio $\eta$, we define the critical value of spin parameter $a_{\rm cri}$ below which the width $\delta r$ of the leftmost $2/3$-resonant island is smaller than $0.001M$. For fixed values of $E$ and $L_z$, the value of $a_{\rm cri}$ varies with $\eta$, and the results of fixing $E=0.9367$ and $L_z=1.7542M$ are shown by the black points for some chosen $\eta$ in Figure~\ref{fig:criticalspin}, which fit very well with a linear function (blue dashed line). From this figure, we find that $a_{\rm cri}$ decreases with increasing symmetric mass ratio $\eta$, meaning that the strength of chaos increases when the values of $a$ and $\eta$ increase. In this figure, we also show a gray curve on which the rotation numbers on the left edge of CZV are $2/3$. Therefore, below the gray curve, the $2/3$-resonant islands do not appear within CZV. In addition, we show the extremality bound $a_{\rm ext}(\eta)$ using the purple curve. Although in Figure~\ref{fig:criticalspin}, we only focus on the parameter space in which the extremality bound is violated, we should emphasize that the chaotic signatures such as Birkhoff islands and plateaus in rotation curves also appear when the extremality bound is satisfied. As we have shown in Figures~\ref{fig:islandshorizon} and \ref{fig:rotationnbhorizon}, upon another choice of $E$ and $L_z$, chaotic signatures also appear when each binary is below the Kerr bound, i.e., $a<(1-2\eta)M$. The islands found there may just be too thin to be detectable in the real gravitational wave events. We consider the configuration that violates the extremality bound in Figure~\ref{fig:criticalspin} mainly due to the fact that the island structures are easier to identify in this configuration. Furthermore, the critical spin $a_{\rm cri}$ for a given $\eta$ depends on the values of $E$ and $L_z$ as well. How $a_{\rm cri}$ would be altered when the parameters $\eta$, $E$, and $L_z$ are all allowed to change is not a trivial question. The main conclusion we would like to draw from Figure~\ref{fig:criticalspin} is that the chaotic behaviors of the system, when fixing $E$ and $L_z$, are enhanced with $a$ and $\eta$. This conclusion is expected to hold also for the configuration where $a<(1-2\eta)M$, and can actually be understood from the observation that the term that breaks the integrability in Eq.~\eqref{21constraint} would also be enhanced by $a$ and $\eta$. In addition, we note that for some cases such as the $1/2$-resonance in Figure \ref{fig:horizonisland} and $2/3$-resonance in Figure \ref{fig:sub21}, the Birkhoff islands appear near the boundary of CZV. Especially, the leftmost boundary of CZV and some Birkhoff islands could be close to the (would-be) event horizons. By tuning $E$, $L_z$ and $a$, the closed CZV may turn into the throat-like one as shown in Figure \ref{fig:czv2}. This can lead to the formation/destruction of the would-be islands by the retreat/intrusion of the event horizon. We indeed do find such cases, which show the interesting effect of the event horizon on the possibility of forming/destroying the resonant islands.

In any case, our results suggest that, the chaotic behaviors of the system would be enhanced by $a$ and $\eta$, and could be too small to measure when these parameters are small or even moderate. A possible implication of these results is that the attempt to detect chaotic signatures of binary systems through the gravitational waves emitted by compact binary coalescence (CBC) of comparable masses, or extreme-mass-ratio-inspiral (EMRI) and intermediate-mass-ratio-inspiral (IMRI) would be extremely difficult.  Of course, whether this is a no-go, i.e., whether the islands can generate observable signatures in gravitational waves, also depends on the dynamical time scale of the system, the population of the events, and the resolution power of the detector.  Typically, the dynamical time scale of systems with tiny $\eta$ such as EMRIs and IMRIs is much longer than that of CBC with sizable $\eta$, so that the passage through islands is detectable, even though the islands are very thin. However, in these cases $a_{\rm cri}$ is in the ultra-spinning regime, for which the events are rare. On the other hand, for the CBC of comparable masses the dynamical time scale is too short to resolve the chaotic signature in the observational data unless for the detectors of high resolution. Due to the implication of our results, it requires detailed study to estimate the chance of detecting chaotic signatures in the gravitational waves of real events. We will come back to this interesting issue elsewhere.

\subsection{Some preliminary 3PN results}\label{3PN}
In this subsection, we investigate the integrability of the effective metric of EOB at 3PN order by examining the phase orbits on Poincar\'e surfaces of section. At 3PN order, two more terms are added into functions $F(r)$ and $G(r)$ written for up to 2PN in \eq{FGH}. We denote these 3PN order functions as $F_{3PN}(r)$ and $G_{3PN}(r)$: 
\bea 
F_{3PN}(r)&=&\frac{2M^3}{r}+\frac{M^4}{r^2}\left(\frac{94}{3}-\frac{41}{32}\pi^2\right)\,, 
\nn \\
 G_{3PN}(r)&=&\frac{1}{\eta} \ln\left[1+6\eta \frac{M^2}{r^2}+2\left(26-3\eta\right)\eta \frac{M^3}{r^3}\right]\,. \label{FGH_3PN}
\eea 
The remaining structure of the effective metric at 3PN is same as the one given in \eq{EOB_meteric_1}. In EOB formulation, at 3PN, the effective Hamiltonian  is not the simple freely moving particle's Hamiltonian like the one written in \eq{H_eff}. Especially, there is an extra term, quadratic in momenta, in the square root in $H_{\rm eff}$ of \eq{H_eff} \cite{Barausse:2009xi,Barausse:2011ys}. As a preliminary study, we ignore this extra term, and continue with the same effective Hamiltonian as in \eq{H_eff} but with 3PN order $F_{3PN}$ and $G_{3PN}$ in the effective metric.

We investigate the phase orbits on the Poincar\'e surface of section for 3PN case with $E=0.942$, $L_z=2.87M$, $a=0.78M$, and $\eta=0.01$, and show the results in Figure \ref{fig:3PN_island_rotnum}. In this case, the EOB metric has a horizon, which is shown by the magenta line in Figure \ref{fig:island_3PN}. We find the $2/3$-resonant islands (blue), even when the symmetric mass ratio is small ($\eta=0.01$). The rotation curve drawn along a horizontal line that crosses the right branch of the islands is shown in Figure~\ref{fig:rotation_number_3PN}. The plateau with $\nu_\theta=2/3$ can be clearly identified. According to this preliminary results, this ``partial" 3PN dynamics may also be chaotic. In addition, the 3PN corrections seem allow us to find Birkhoff islands even when $\eta$ is small.

\begin{figure}
\centering
\begin{subfigure}{.5\textwidth}
  \centering
  \includegraphics[width=0.85\linewidth]{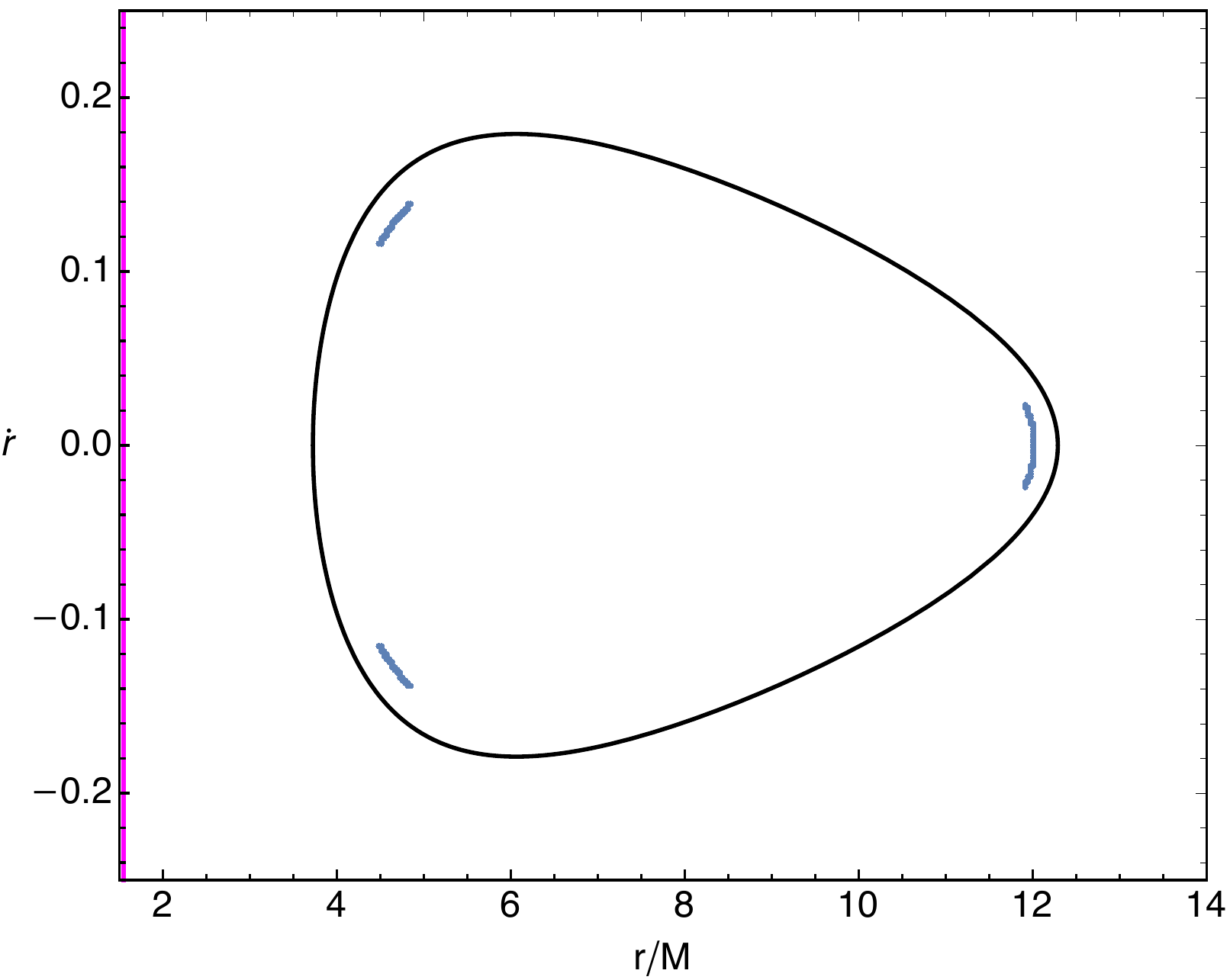}
  \caption{}
  \label{fig:island_3PN}
\end{subfigure}%
\begin{subfigure}{.5\textwidth}
  \centering
  \includegraphics[width=0.85\linewidth]{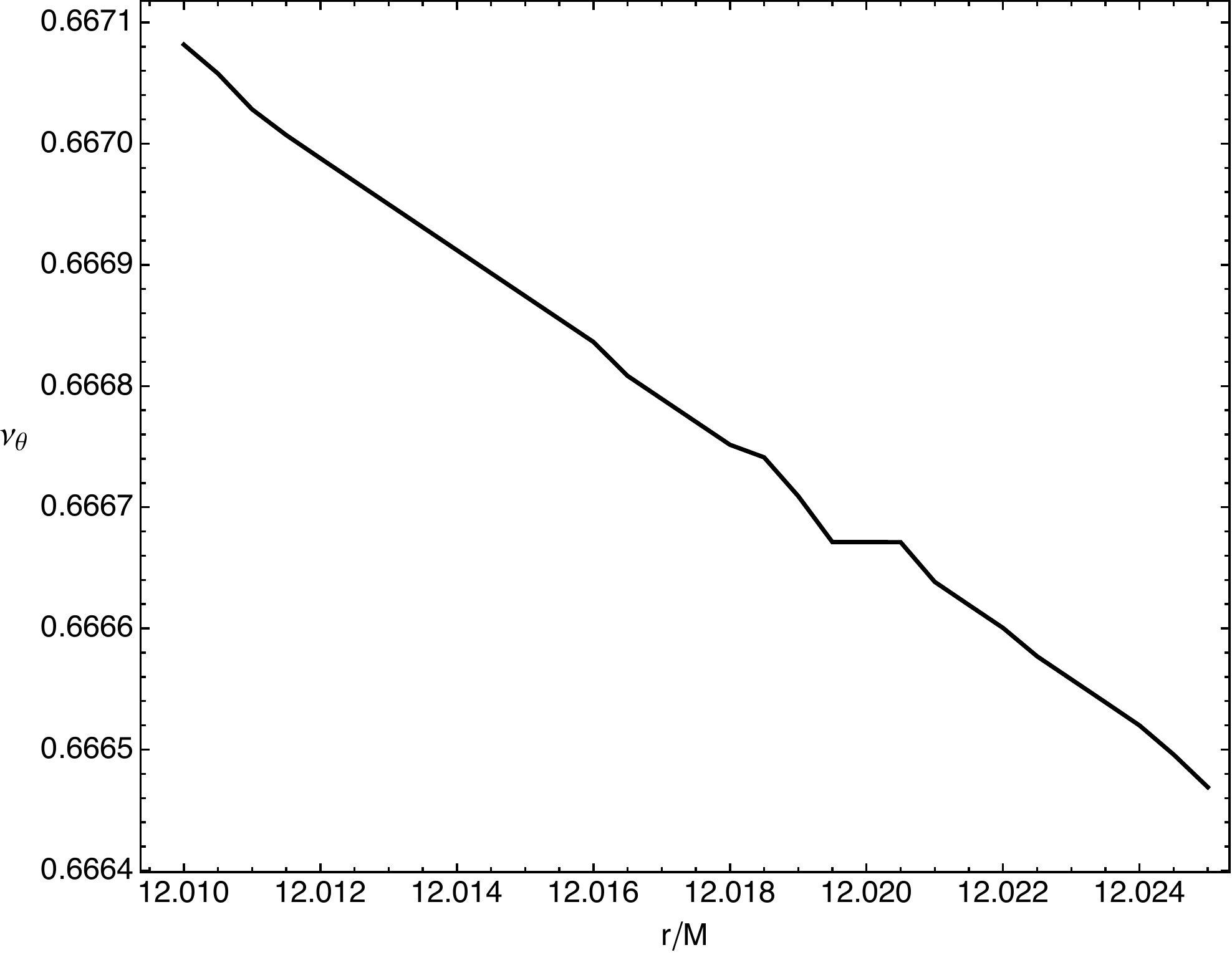}
  \caption{}
  \label{fig:rotation_number_3PN}
\end{subfigure}
\caption{(a) The $2/3$-Birkhoff islands for the 3PN effective metric for $a=0.78M$ and $\eta=0.01$. The initial conditions for the islands are $E=0.942$, $L_z=2.87M$, and $(r_0, y_0, \dot{r}_0)=(12.02M, 0, 0)$. The magenta line indicates the event horizon. (b) The rotation curve drawn along a horizontal line that crosses the rightmost branch of the islands. The plateau with $\nu_\theta=2/3$ can be identified.}
\label{fig:3PN_island_rotnum}
\end{figure}

\section{Conclusion}\label{sec.conclude}
The inspiral dynamics of the compact binary black holes is intrinsically non-linear and non-integrable due to the non-linearity of general relativity, which manifests as the higher-orders of post-Newtonian corrections.  Thus, it is very interesting to study the possibility of detecting such non-integrability and the associated chaotic behaviors through the astronomical observations such as the gravitational wave detection. If such chaotic signatures exist, one will expect to detect them in the future with more advanced technology. 

In this work, we consider the effective-one-body (EOB) formulation of the inspiral dynamics, and study the phase orbits to find out the Birkhoff chains of islands, which encode the chaotic signatures. The advantage of the EOB formulation than the usual binary dynamics is the reduction of the dynamical variables into the ones of test-body moving in a blackhole-like effective metric. The reduction holds whenever the EOB dynamics is a good approximation to the original binary dynamics.  This makes easier for the formal check of the non-integrability and the numerical study of the chaotic behavior. As a preliminary study, we consider EOB dynamics up to 2PN order and some partial 3PN order terms. This is the lowest PN order for inspiral dynamics to be non-integrable. For simplicity, we also turn off the spin of the test-body. As expected, our results show the clear signatures of non-integrability in the form of the Birkhoff islands. Moreover, we also show that the chaotic behaviors would be enhanced when the spin parameter $a$ and the symmetric mass ratio $\eta$ increase. This yield some implications for the issue of detecting chaos in the real gravitational wave events. 

In summary, this work provides a new methodology to study the chaotic behaviors of inspiral dynamics of binary compact objects, and exhibits the possible chaotic signatures. It is worthy to extend our method to higher PN EOB dynamics and include spin of the test-body to obtain a more complete picture of the chaos in such dynamical systems and examine more closely on the issue of their detectability.

\acknowledgments
The authors thank Georgios Lukes-Gerakopoulos, Alejandro C\'ardenas-Avenda\~no, Hsu-Wen Chiang, and Kyriakos Destounis for the fruitful suggestions on the technical parts of this work. FLL is supported by Taiwan's Ministry of Science and Technology (MoST) through Grant No.~109-2112-M-003-007-MY3. AP is supported by MoST grant No. 110-2811-M-003-507-MY2.

\appendix

\section{Summary of deriving EOB mapping  and comment on the integrability}\label{app_A}

In this appendix, we would like to summarize the procedure of constructing the EOB mapping given in \cite{Damour:2001tu,Barausse:2009xi,Barausse:2011ys}, and also take the chance to comment on the integrability of the seemingly equivalent PN dynamics such as the one given in \cite{Konigsdorffer:2005sc}.

Following \cite{Damour:2001tu}, the authors of \cite{Barausse:2009xi,Barausse:2011ys} gave a clear procedure to construct the EOB Hamiltonian for the dynamics of binary spinning black holes. They can be summarized as follows:

\begin{enumerate}[(i)]
\item  Start with PN-expanded ADM Hamiltonian and then perform a canonical transformation (Lie method) to PN-expanded Hamiltonian in EOB coordinates. 

\item  Use the results of (i) to compute the corresponding PN-expanded EOB Hamiltonian.

\item  PN-expand the deformed-Kerr Hamiltonian for a spinning test particle.

\item  Compare (ii) and (iii) to obtain the mapping between the spin variables in the real and effective descriptions.
\end{enumerate}

In general, the effective Hamiltonian of the EOB dynamics should formally be 
\be 
H_{\rm eff}= H^{NS} + H^S
\ee 
where $NS$ denotes the part that does not involve spin of the test particle, ${\bf S}^*$, and $S$ the part that involves ${\bf S}^*$. The PN-expanded form of $H^{NS}=\beta^i p_i+\alpha \sqrt{m^2+\gamma^{ij}p_i p_j+ Q_A(p)}$ can be summarized as follows \footnote{$Q_A(p)$ contains the non-geometric terms starting from 3PN order. Therefore, there is no such term in the 2PN order considered in sections \ref{EOB_rev} and \ref{2PN_chaos}, and in the preliminary study of the 3PN order in section \ref{3PN} this term is neglected for simplicity.}:
\be
H^{NS}=H^{NS}_{S=0}+ H^{NS}_{SO,1.5PN} + H^{NS}_{SS, 2PN}\,.
\ee 
Here $H^{NS}_{S=0}$ is the PN-Hamiltonian for the non-spinning binary, and the last two terms involve only the spin of the deformed-Kerr denoted by ${\bf S}_{\rm Kerr}$, which is different from the spin of the test particle denoted by ${\bf S}^*$. The $H^{NS}_{SO,1.5PN}$ is the 1.5PN spin-orbital interaction which is linear in ${\bf S}_{\rm Kerr}$, and $H^{NS}_{SS, 2PN}$ contains the spin-spin interactions which are quadratic in ${\bf S}_{\rm Kerr}$.  Note that $H^{NS}$ is controlled by two metric functions $A(r)$ and $D(r)$, which are the same as the ones for the non-spinning binary up to 3PN. The terms involving ${\bf S}_{\rm Kerr}$ come from the Kerr terms and the additional $\tilde{\omega}_{fd}$, all of which vanish if ${\bf S}_{\rm Kerr}$ is set to zero. 

On the other hand, $H^S$ mainly contains terms that involve ${\bf S}^*$ so it vanishes if ${\bf S}^*$ is set to zero. The PN-expanded form of $H^S$ can be summarized as follows:
\be
H^S= H^S_{SO, 1.5PN} + H^S_{SO, 2.5PN} + H^S_{SS, 2PN}\,.
\ee 
The $H^S_{SO, 1.5PN}$ is the 1.5PN spin-orbital interaction that is linear in ${\bf S}^*$. The $H^S_{SO, 2.5PN}$ is the 2.5PN spin-orbital interactions linear in ${\bf S}^*$ but also involved ${\bf S}_{\rm Kerr}$.  The last term $H^S_{SS, 2PN}$ is spin-spin interaction and is quadratic in ${\bf S}^*$. This term is put in by hand since the Hamiltonian of the test particle is valid only at the linear order of ${\bf S}^*$. 

Step (iv) is to obtain the map below between ${\bf S}_{\rm Kerr}$, ${\bf S}^*$ of the EOB effective Hamiltonian and ${\bf S}_1$ and ${\bf S}_2$ of the ADM Hamiltonian. Instead of using ${\bf S}_1$ and ${\bf S}_2$, one introduces 
\be 
\bm{\sigma}={\bf S}_1+{\bf S}_2\,, \qquad \bm{\sigma}^*={\bf S}_1 {m_2\over m_1}+ {\bf S}_2 {m_1\over m_2}\,.
\ee 
The map obtained in (iv) is the following
\be 
{\bf S}_{\rm Kerr}=\bm{\sigma} + {1\over c^2} \bm{\Delta}_{\sigma}\,, \qquad {\bf S}^*=\bm{\sigma}^* + {1\over c^2}\bm{\Delta}_{\sigma^*} 
\ee 
where 
\be 
\bm{\Delta}_{\sigma}=-{1\over 16} \left\{ 12 \bm{\Delta}_{\sigma^*} + \eta \Big[ {2 M\over r} ( 4\bm{\sigma} - 7\bm{\sigma}^*) + 6 (\hat{\bm{p}}\cdot \hat{\bm{n}})^2 ( 6 \bm{\sigma} + 5 \bm{\sigma}^*) - \hat{p}^2 (3\bm{\sigma} + 4\bm{\sigma}^*)\Big] \right\}
\ee
but ${\bf \Delta}_{\sigma^*}$ is an arbitrary function going to zero at least linearly in $\eta$ when $\eta\rightarrow 0$ to get correct test-particle limit. Here $\hat{\bm{n}}={\bf r}/r$ and $\hat{\bm{p}}={\bf p}/m$. In \cite{Barausse:2009xi}, they chose $\bm{\Delta}_{\sigma^*}$ so that $\bm{\Delta}_{\sigma}=0$. In contrast, for our purpose in this work, we choose $\bm{ \Delta}_{\sigma^*}=-c^2 \bm{\sigma}^*$ so that ${\bf S}^*=0$.  Since $|{\bf S}_1|$, $|{\bf S}_2|$ and ${\bf L}+{\bf S}_1+{\bf S}_2$ are conserved but not $\bm{\sigma}$ or $\bm{\sigma}^*$. Thus, both the choice of \cite{Barausse:2009xi} and ours are dynamical constraints. Furthermore, in \cite{Barausse:2011ys} a new EOB frame is adopted so that the EOB spins ${\bf S}_{\rm Kerr}=\bm{\sigma}$ and ${\bf S}^*$ are independent of dynamical variables other than ${\bf S}_{1,2}$. This new EOB frame is related to the above one by a canonical transformation and a redefinition of the Hamilton-Jacobi function, and the corresponding EOB effective Hamiltonians differ at 3PN order. Again, in this new frame, we can choose to set ${\bf S}^*$ to zero as a dynamical constraint.

Finally, we comment on the integrable example considered in \cite{Konigsdorffer:2005sc} and compare with the above EOB case. The ADM Hamiltonian adopted in \cite{Konigsdorffer:2005sc} is formally the following
\be\label{H_Gopa}
H_{Gopa}= H^{NS}_{S=0,3PN} + H^{S}_{SO}
\ee
where $H^{NS}_{S=0,3PN}$ is the 3PN Hamiltonian of non-spinning binary, which should correspond to EOB effective Hamiltonian of a deformed Schwarzschild background; and $H^{S}_{SO}$ is the spin-orbital Hamiltonian linear in the following effective spin
\be 
{\bf S}_{\rm eff}= g_1 {\bf S}_1 + g_2 {\bf S}_2
\ee
where the gyromagnetic factors are
\bea 
g_1 &=& 2\eta \Big(1+ {3 m_2 \over 4m_1}\Big)\,,
\\
g_2 &=& 2\eta \Big(1+ {3 m_1 \over 4m_2}\Big)\,.
\eea
The authors of \cite{Konigsdorffer:2005sc} found integrable solutions for two special cases: (1) $m_1=m_2$, and (2) either ${\bf S}_1$ or ${\bf S}_2$ vanishes. The Hamiltonian used in \cite{Konigsdorffer:2005sc} differs from the ADM Hamiltonian adopted in \cite{Barausse:2009xi,Barausse:2011ys} where gyromagnetic factors $g_i$ of latter involve dynamical variables $\hat{\bm{p}'}^2$, $\hat{\bm{p}'}\cdot \hat{\bm{n}'}$ and $M/r$ in addition to the symmetric mass ratio $\eta$ and the mass ratio $m_1/m_2$. Note that the prime is used to remind that the corresponding quantity is associated with the ADM coordinate to distinguish from those in the non-prime EOB coordinate. Also, in  \cite{Barausse:2009xi,Barausse:2011ys} there also includes the 2PN spin-spin Hamiltonian
\be
H^{ADM}_{SS}={1\over c^4} {\eta \over 2 r^3}[3 (\hat{\bm{n}'}\cdot \bm{\sigma}_0)^2 -\bm{ \sigma}_0^2]
\ee
with $\bm{\sigma}_0=\bm{\sigma} + \bm{\sigma}_*$. This term could be responsible for the quadratic terms of ${\bf S}_{\rm Kerr}$. Thus, even if we do not work out the EOB Hamiltonian from $H_{Gopa}$ of \eq{H_Gopa} by canonical transformation, we know that the resultant EOB effective Hamiltonian will be different from the one obtained in \cite{Barausse:2009xi,Barausse:2011ys}. Thus, the integrable solutions in \cite{Konigsdorffer:2005sc} are not in contradiction to the non-integrability and chaotic features found in this work because the adopted ADM Hamiltonians are different. 

\bibliographystyle{jhep}
\bibliography{references.bib} % your references

\providecommand{\href}[2]{#2}\begingroup\raggedright\begin{thebibliography}{10}

\bibitem{LIGOScientific:2021djp}
{\scshape LIGO Scientific, VIRGO, KAGRA} collaboration, \emph{{GWTC-3: Compact
  Binary Coalescences Observed by LIGO and Virgo During the Second Part of the
  Third Observing Run}},  \href{https://arxiv.org/abs/2111.03606}{{\ttfamily
  2111.03606}}.

\bibitem{Buonanno:1998gg}
A.~Buonanno and T.~Damour, \emph{{Effective one-body approach to general
  relativistic two-body dynamics}},
  \href{https://doi.org/10.1103/PhysRevD.59.084006}{\emph{Phys. Rev. D}
  {\bfseries 59} (1999) 084006}
  [\href{https://arxiv.org/abs/gr-qc/9811091}{{\ttfamily gr-qc/9811091}}].

\bibitem{Damour:2000we}
T.~Damour, P.~Jaranowski and G.~Schaefer, \emph{{On the determination of the
  last stable orbit for circular general relativistic binaries at the third
  postNewtonian approximation}},
  \href{https://doi.org/10.1103/PhysRevD.62.084011}{\emph{Phys. Rev. D}
  {\bfseries 62} (2000) 084011}
  [\href{https://arxiv.org/abs/gr-qc/0005034}{{\ttfamily gr-qc/0005034}}].

\bibitem{Damour:2001tu}
T.~Damour, \emph{{Coalescence of two spinning black holes: an effective
  one-body approach}},
  \href{https://doi.org/10.1103/PhysRevD.64.124013}{\emph{Phys. Rev. D}
  {\bfseries 64} (2001) 124013}
  [\href{https://arxiv.org/abs/gr-qc/0103018}{{\ttfamily gr-qc/0103018}}].

\bibitem{Barausse:2009xi}
E.~Barausse and A.~Buonanno, \emph{{An Improved effective-one-body Hamiltonian
  for spinning black-hole binaries}},
  \href{https://doi.org/10.1103/PhysRevD.81.084024}{\emph{Phys. Rev. D}
  {\bfseries 81} (2010) 084024}
  [\href{https://arxiv.org/abs/0912.3517}{{\ttfamily 0912.3517}}].

\bibitem{Barausse:2011ys}
E.~Barausse and A.~Buonanno, \emph{{Extending the effective-one-body
  Hamiltonian of black-hole binaries to include next-to-next-to-leading
  spin-orbit couplings}},
  \href{https://doi.org/10.1103/PhysRevD.84.104027}{\emph{Phys. Rev. D}
  {\bfseries 84} (2011) 104027}
  [\href{https://arxiv.org/abs/1107.2904}{{\ttfamily 1107.2904}}].

\bibitem{Cornish:2001jy}
N.J.~Cornish, \emph{{Chaos and gravitational waves}},
  \href{https://doi.org/10.1103/PhysRevD.64.084011}{\emph{Phys. Rev. D}
  {\bfseries 64} (2001) 084011}
  [\href{https://arxiv.org/abs/gr-qc/0106062}{{\ttfamily gr-qc/0106062}}].

\bibitem{Levin:1999zx}
J.J.~Levin, \emph{{Gravity waves, chaos, and spinning compact binaries}},
  \href{https://doi.org/10.1103/PhysRevLett.84.3515}{\emph{Phys. Rev. Lett.}
  {\bfseries 84} (2000) 3515}
  [\href{https://arxiv.org/abs/gr-qc/9910040}{{\ttfamily gr-qc/9910040}}].

\bibitem{Levin:2000md}
J.J.~Levin, \emph{{The fate of chaotic binaries}},
  \href{https://doi.org/10.1103/PhysRevD.67.044013}{\emph{Phys. Rev. D}
  {\bfseries 67} (2003) 044013}
  [\href{https://arxiv.org/abs/gr-qc/0010100}{{\ttfamily gr-qc/0010100}}].

\bibitem{Hughes:2000nzm}
S.A.~Hughes, \emph{{Comment on `Gravity waves, chaos, and spinning compact
  binaries'}}, \href{https://doi.org/10.1103/PhysRevLett.85.5480}{\emph{Phys.
  Rev. Lett.} {\bfseries 85} (2000) 5480}
  [\href{https://arxiv.org/abs/gr-qc/0101024}{{\ttfamily gr-qc/0101024}}].

\bibitem{Cornish:2003ig}
N.J.~Cornish and J.J.~Levin, \emph{{Lyapunov timescales and black hole
  binaries}}, \href{https://doi.org/10.1088/0264-9381/20/9/304}{\emph{Class.
  Quant. Grav.} {\bfseries 20} (2003) 1649}
  [\href{https://arxiv.org/abs/gr-qc/0304056}{{\ttfamily gr-qc/0304056}}].

\bibitem{Schnittman:2001mz}
J.D.~Schnittman and F.A.~Rasio, \emph{{Ruling out chaos in compact binary
  systems}}, \href{https://doi.org/10.1103/PhysRevLett.87.121101}{\emph{Phys.
  Rev. Lett.} {\bfseries 87} (2001) 121101}
  [\href{https://arxiv.org/abs/gr-qc/0107082}{{\ttfamily gr-qc/0107082}}].

\bibitem{Cornish:2002gz}
N.J.~Cornish and J.J.~Levin, \emph{{Comment on `Ruling out chaos in compact
  binary systems'}},
  \href{https://doi.org/10.1103/PhysRevLett.89.179001}{\emph{Phys. Rev. Lett.}
  {\bfseries 89} (2002) 179001}
  [\href{https://arxiv.org/abs/gr-qc/0207020}{{\ttfamily gr-qc/0207020}}].

\bibitem{Hartl:2004xr}
M.D.~Hartl and A.~Buonanno, \emph{{The Dynamics of precessing binary black
  holes using the post-Newtonian approximation}},
  \href{https://doi.org/10.1103/PhysRevD.71.024027}{\emph{Phys. Rev. D}
  {\bfseries 71} (2005) 024027}
  [\href{https://arxiv.org/abs/gr-qc/0407091}{{\ttfamily gr-qc/0407091}}].

\bibitem{Wu:2007zze}
X.~Wu and Y.~Xie, \emph{{Revisit on 'Ruling out chaos in compact binary
  systems'}}, \href{https://doi.org/10.1103/PhysRevD.76.124004}{\emph{Phys.
  Rev. D} {\bfseries 76} (2007) 124004}
  [\href{https://arxiv.org/abs/1004.5057}{{\ttfamily 1004.5057}}].

\bibitem{Wu:2010mv}
X.~Wu and Y.~Xie, \emph{{Symplectic structure of post-Newtonian Hamiltonian for
  spinning compact binaries}},
  \href{https://doi.org/10.1103/PhysRevD.81.084045}{\emph{Phys. Rev. D}
  {\bfseries 81} (2010) 084045}
  [\href{https://arxiv.org/abs/1004.4549}{{\ttfamily 1004.4549}}].

\bibitem{2011GReGr..43.2185W}
X.~{Wu} and S.-Y.~{Zhong}, \emph{{Regular dynamics of canonical post-Newtonian
  Hamiltonian for spinning compact binaries with next-to-leading order
  spin-orbit interactions}},
  \href{https://doi.org/10.1007/s10714-011-1171-0}{\emph{General Relativity and
  Gravitation} {\bfseries 43} (2011) 2185}.

\bibitem{Cho:2019brd}
G.~Cho and H.M.~Lee, \emph{{Analytic Keplerian-type parametrization for general
  spinning compact binaries with leading order spin-orbit interactions}},
  \href{https://doi.org/10.1103/PhysRevD.100.044046}{\emph{Phys. Rev. D}
  {\bfseries 100} (2019) 044046}
  [\href{https://arxiv.org/abs/1908.02927}{{\ttfamily 1908.02927}}].

\bibitem{Tanay:2020gfb}
S.~Tanay, L.C.~Stein and J.T.~G\'alvez~Ghersi, \emph{{Integrability of
  eccentric, spinning black hole binaries up to second post-Newtonian order}},
  \href{https://doi.org/10.1103/PhysRevD.103.064066}{\emph{Phys. Rev. D}
  {\bfseries 103} (2021) 064066}
  [\href{https://arxiv.org/abs/2012.06586}{{\ttfamily 2012.06586}}].

\bibitem{Tanay:2021bff}
S.~Tanay, G.~Cho and L.C.~Stein, \emph{{Action-angle variables of a binary
  black-hole with arbitrary eccentricity, spins, and masses at 1.5
  post-Newtonian order}},  \href{https://arxiv.org/abs/2110.15351}{{\ttfamily
  2110.15351}}.

\bibitem{Morras:2021atg}
G.~Morr\'as, J.~Garc\'\i{}a-Bellido and S.~Nesseris, \emph{{Search for black
  hole hyperbolic encounters with gravitational wave detectors}},
  \href{https://doi.org/10.1016/j.dark.2021.100932}{\emph{Phys. Dark Univ.}
  {\bfseries 35} (2022) 100932}
  [\href{https://arxiv.org/abs/2110.08000}{{\ttfamily 2110.08000}}].

\bibitem{Gopakumar:2005zz}
A.~Gopakumar and C.~Konigsdorffer, \emph{{The Deterministic nature of
  conservative post-Newtonian accurate dynamics of compact binaries with
  leading order spin-orbit interaction}},
  \href{https://doi.org/10.1103/PhysRevD.72.121501}{\emph{Phys. Rev. D}
  {\bfseries 72} (2005) 121501}
  [\href{https://arxiv.org/abs/gr-qc/0511009}{{\ttfamily gr-qc/0511009}}].

\bibitem{Konigsdorffer:2005sc}
C.~Konigsdorffer and A.~Gopakumar, \emph{{Post-Newtonian accurate parametric
  solution to the dynamics of spinning compact binaries in eccentric orbits:
  The Leading order spin-orbit interaction}},
  \href{https://doi.org/10.1103/PhysRevD.71.024039}{\emph{Phys. Rev. D}
  {\bfseries 71} (2005) 024039}
  [\href{https://arxiv.org/abs/gr-qc/0501011}{{\ttfamily gr-qc/0501011}}].

\bibitem{Huang:2014ska}
G.~Huang, X.~Ni and X.~Wu, \emph{{Chaos in two black holes with next-to-leading
  order spin-spin interactions}},
  \href{https://doi.org/10.1140/epjc/s10052-014-3012-2}{\emph{Eur. Phys. J. C}
  {\bfseries 74} (2014) 3012}
  [\href{https://arxiv.org/abs/1403.0378}{{\ttfamily 1403.0378}}].

\bibitem{Wu:2015cqa}
X.~Wu and G.~Huang, \emph{{Ruling out chaos in comparable mass compact binary
  systems with one body spinning}},
  \href{https://doi.org/10.1093/mnras/stv1485}{\emph{Mon. Not. Roy. Astron.
  Soc.} {\bfseries 452} (2015) 3167}.

\bibitem{Huang:2016vfk}
L.~Huang and X.~Wu, \emph{{Second post-Newtonian Lagrangian dynamics of
  spinning compact binaries}},
  \href{https://doi.org/10.1140/epjc/s10052-016-4339-7}{\emph{Eur. Phys. J. C}
  {\bfseries 76} (2016) 488}
  [\href{https://arxiv.org/abs/1604.05810}{{\ttfamily 1604.05810}}].

\bibitem{arnold1989mathematical}
V.~Arnold, \emph{Mathematical methods of classical mechanics}, vol.~60,
  Springer (1989).

\bibitem{Frolov:2017kze}
V.~Frolov, P.~Krtous and D.~Kubiznak, \emph{{Black holes, hidden symmetries,
  and complete integrability}},
  \href{https://doi.org/10.1007/s41114-017-0009-9}{\emph{Living Rev. Rel.}
  {\bfseries 20} (2017) 6} [\href{https://arxiv.org/abs/1705.05482}{{\ttfamily
  1705.05482}}].

\bibitem{Papadopoulos:2018nvd}
G.O.~Papadopoulos and K.D.~Kokkotas, \emph{{Preserving Kerr symmetries in
  deformed spacetimes}},
  \href{https://doi.org/10.1088/1361-6382/aad7f4}{\emph{Class. Quant. Grav.}
  {\bfseries 35} (2018) 185014}
  [\href{https://arxiv.org/abs/1807.08594}{{\ttfamily 1807.08594}}].

\bibitem{Compere:2021kjz}
G.~Comp\`ere and A.~Druart, \emph{{Complete set of quasi-conserved quantities
  for spinning particles around Kerr}},
  \href{https://doi.org/10.21468/SciPostPhys.12.1.012}{\emph{SciPost Phys.}
  {\bfseries 12} (2022) 012}
  [\href{https://arxiv.org/abs/2105.12454}{{\ttfamily 2105.12454}}].

\bibitem{2002ocda.book.....C}
G.~{Contopoulos}, \emph{{Order and chaos in dynamical astronomy}} (2002).

\bibitem{Lukes-Gerakopoulos:2016bup}
G.~Lukes-Gerakopoulos, M.~Katsanikas, P.A.~Patsis and J.~Seyrich, \emph{{The
  dynamics of a spinning particle in a linear in spin Hamiltonian
  approximation}},
  \href{https://doi.org/10.1103/PhysRevD.94.024024}{\emph{Phys. Rev. D}
  {\bfseries 94} (2016) 024024}
  [\href{https://arxiv.org/abs/1606.09171}{{\ttfamily 1606.09171}}].

\bibitem{Harms:2016ctx}
E.~Harms, G.~Lukes-Gerakopoulos, S.~Bernuzzi and A.~Nagar, \emph{{Spinning test
  body orbiting around a Schwarzschild black hole: Circular dynamics and
  gravitational-wave fluxes}},
  \href{https://doi.org/10.1103/PhysRevD.94.104010}{\emph{Phys. Rev. D}
  {\bfseries 94} (2016) 104010}
  [\href{https://arxiv.org/abs/1609.00356}{{\ttfamily 1609.00356}}].

\bibitem{Lukes-Gerakopoulos:2016udm}
G.~Lukes-Gerakopoulos, \emph{{Spinning particles moving around black holes:
  integrability and chaos}},  in \emph{{14th Marcel Grossmann Meeting on Recent
  Developments in Theoretical and Experimental General Relativity,
  Astrophysics, and Relativistic Field Theories}}, vol.~2, pp.~1960--1965,
  2017, \href{https://doi.org/10.1142/9789813226609_0209}{DOI}
  [\href{https://arxiv.org/abs/1606.09430}{{\ttfamily 1606.09430}}].

\bibitem{Lukes-Gerakopoulos:2017vkj}
G.~Lukes-Gerakopoulos, E.~Harms, S.~Bernuzzi and A.~Nagar, \emph{{Spinning
  test-body orbiting around a Kerr black hole: circular dynamics and
  gravitational-wave fluxes}},
  \href{https://doi.org/10.1103/PhysRevD.96.064051}{\emph{Phys. Rev. D}
  {\bfseries 96} (2017) 064051}
  [\href{https://arxiv.org/abs/1707.07537}{{\ttfamily 1707.07537}}].

\bibitem{Zelenka:2019nyp}
O.~Zelenka, G.~Lukes-Gerakopoulos, V.~Witzany and O.~Kop\'a\v{c}ek,
  \emph{{Growth of resonances and chaos for a spinning test particle in the
  Schwarzschild background}},
  \href{https://doi.org/10.1103/PhysRevD.101.024037}{\emph{Phys. Rev. D}
  {\bfseries 101} (2020) 024037}
  [\href{https://arxiv.org/abs/1911.00414}{{\ttfamily 1911.00414}}].

\bibitem{Skoupy:2021asz}
V.~Skoup\'y and G.~Lukes-Gerakopoulos, \emph{{Spinning test body orbiting
  around a Kerr black hole: Eccentric equatorial orbits and their asymptotic
  gravitational-wave fluxes}},
  \href{https://doi.org/10.1103/PhysRevD.103.104045}{\emph{Phys. Rev. D}
  {\bfseries 103} (2021) 104045}
  [\href{https://arxiv.org/abs/2102.04819}{{\ttfamily 2102.04819}}].

\bibitem{Timogiannis:2021ung}
I.~Timogiannis, G.~Lukes-Gerakopoulos and T.A.~Apostolatos, \emph{{Spinning
  test body orbiting around a Schwarzschild black hole: Comparing spin
  supplementary conditions for circular equatorial orbits}},
  \href{https://doi.org/10.1103/PhysRevD.104.024042}{\emph{Phys. Rev. D}
  {\bfseries 104} (2021) 024042}
  [\href{https://arxiv.org/abs/2104.11183}{{\ttfamily 2104.11183}}].

\bibitem{Skoupy:2022adh}
V.~Skoup\'y and G.~Lukes-Gerakopoulos, \emph{{Adiabatic equatorial inspirals of
  a spinning body into a Kerr black hole}},
  \href{https://doi.org/10.1103/PhysRevD.105.084033}{\emph{Phys. Rev. D}
  {\bfseries 105} (2022) 084033}
  [\href{https://arxiv.org/abs/2201.07044}{{\ttfamily 2201.07044}}].

\bibitem{Timogiannis:2022bks}
I.~Timogiannis, G.~Lukes-Gerakopoulos and T.A.~Apostolatos, \emph{{Spinning
  test body orbiting around a Kerr black hole: Comparing spin supplementary
  conditions for circular equatorial orbits}},
  \href{https://doi.org/10.1103/PhysRevD.106.044039}{\emph{Phys. Rev. D}
  {\bfseries 106} (2022) 044039}
  [\href{https://arxiv.org/abs/2206.11149}{{\ttfamily 2206.11149}}].

\bibitem{Suzuki:1996gm}
S.~Suzuki and K.-i.~Maeda, \emph{{Chaos in Schwarzschild space-time: The motion
  of a spinning particle}},
  \href{https://doi.org/10.1103/PhysRevD.55.4848}{\emph{Phys. Rev. D}
  {\bfseries 55} (1997) 4848}
  [\href{https://arxiv.org/abs/gr-qc/9604020}{{\ttfamily gr-qc/9604020}}].

\bibitem{Hartl:2002ig}
M.D.~Hartl, \emph{{Dynamics of spinning test particles in Kerr space-time}},
  \href{https://doi.org/10.1103/PhysRevD.67.024005}{\emph{Phys. Rev. D}
  {\bfseries 67} (2003) 024005}
  [\href{https://arxiv.org/abs/gr-qc/0210042}{{\ttfamily gr-qc/0210042}}].

\bibitem{Hartl:2003da}
M.D.~Hartl, \emph{{A Survey of spinning test particle orbits in Kerr
  space-time}}, \href{https://doi.org/10.1103/PhysRevD.67.104023}{\emph{Phys.
  Rev. D} {\bfseries 67} (2003) 104023}
  [\href{https://arxiv.org/abs/gr-qc/0302103}{{\ttfamily gr-qc/0302103}}].

\bibitem{Kiuchi:2004bv}
K.~Kiuchi and K.-i.~Maeda, \emph{{Gravitational waves from chaotic dynamical
  system}}, \href{https://doi.org/10.1103/PhysRevD.70.064036}{\emph{Phys. Rev.
  D} {\bfseries 70} (2004) 064036}
  [\href{https://arxiv.org/abs/gr-qc/0404124}{{\ttfamily gr-qc/0404124}}].

\bibitem{Gair:2007kr}
J.R.~Gair, C.~Li and I.~Mandel, \emph{{Observable Properties of Orbits in Exact
  Bumpy Spacetimes}},
  \href{https://doi.org/10.1103/PhysRevD.77.024035}{\emph{Phys. Rev. D}
  {\bfseries 77} (2008) 024035}
  [\href{https://arxiv.org/abs/0708.0628}{{\ttfamily 0708.0628}}].

\bibitem{Han:2008zzf}
W.~Han, \emph{{Chaos and dynamics of spinning particles in Kerr spacetime}},
  \href{https://doi.org/10.1007/s10714-007-0598-9}{\emph{Gen. Rel. Grav.}
  {\bfseries 40} (2008) 1831}
  [\href{https://arxiv.org/abs/1006.2229}{{\ttfamily 1006.2229}}].

\bibitem{Flanagan:2010cd}
E.E.~Flanagan and T.~Hinderer, \emph{{Transient resonances in the inspirals of
  point particles into black holes}},
  \href{https://doi.org/10.1103/PhysRevLett.109.071102}{\emph{Phys. Rev. Lett.}
  {\bfseries 109} (2012) 071102}
  [\href{https://arxiv.org/abs/1009.4923}{{\ttfamily 1009.4923}}].

\bibitem{Apostolatos:2009vu}
T.A.~Apostolatos, G.~Lukes-Gerakopoulos and G.~Contopoulos, \emph{{How to
  Observe a Non-Kerr Spacetime Using Gravitational Waves}},
  \href{https://doi.org/10.1103/PhysRevLett.103.111101}{\emph{Phys. Rev. Lett.}
  {\bfseries 103} (2009) 111101}
  [\href{https://arxiv.org/abs/0906.0093}{{\ttfamily 0906.0093}}].

\bibitem{Lukes-Gerakopoulos:2010ipp}
G.~Lukes-Gerakopoulos, T.A.~Apostolatos and G.~Contopoulos, \emph{{Observable
  signature of a background deviating from the Kerr metric}},
  \href{https://doi.org/10.1103/PhysRevD.81.124005}{\emph{Phys. Rev. D}
  {\bfseries 81} (2010) 124005}
  [\href{https://arxiv.org/abs/1003.3120}{{\ttfamily 1003.3120}}].

\bibitem{Contopoulos:2011dz}
G.~Contopoulos, G.~Lukes-Gerakopoulos and T.A.~Apostolatos, \emph{{Orbits in a
  non-Kerr Dynamical System}},
  \href{https://doi.org/10.1142/S0218127411029768}{\emph{Int. J. Bifurc. Chaos}
  {\bfseries 21} (2011) 2261}
  [\href{https://arxiv.org/abs/1108.5057}{{\ttfamily 1108.5057}}].

\bibitem{Brink:2013nna}
J.~Brink, M.~Geyer and T.~Hinderer, \emph{{Orbital resonances around Black
  holes}}, \href{https://doi.org/10.1103/PhysRevLett.114.081102}{\emph{Phys.
  Rev. Lett.} {\bfseries 114} (2015) 081102}
  [\href{https://arxiv.org/abs/1304.0330}{{\ttfamily 1304.0330}}].

\bibitem{Brink:2015roa}
J.~Brink, M.~Geyer and T.~Hinderer, \emph{{Astrophysics of resonant orbits in
  the Kerr metric}},
  \href{https://doi.org/10.1103/PhysRevD.91.083001}{\emph{Phys. Rev. D}
  {\bfseries 91} (2015) 083001}
  [\href{https://arxiv.org/abs/1501.07728}{{\ttfamily 1501.07728}}].

\bibitem{Cardenas-Avendano:2018ocb}
A.~C\'ardenas-Avenda\~no, A.F.~Gutierrez, L.A.~Pach\'on and N.~Yunes,
  \emph{{The exact dynamical Chern\textendash{}Simons metric for a spinning
  black hole possesses a fourth constant of motion: A dynamical-systems-based
  conjecture}}, \href{https://doi.org/10.1088/1361-6382/aad06f}{\emph{Class.
  Quant. Grav.} {\bfseries 35} (2018) 165010}
  [\href{https://arxiv.org/abs/1804.04002}{{\ttfamily 1804.04002}}].

\bibitem{Destounis:2020kss}
K.~Destounis, A.G.~Suvorov and K.D.~Kokkotas, \emph{{Testing spacetime symmetry
  through gravitational waves from extreme-mass-ratio inspirals}},
  \href{https://doi.org/10.1103/PhysRevD.102.064041}{\emph{Phys. Rev. D}
  {\bfseries 102} (2020) 064041}
  [\href{https://arxiv.org/abs/2009.00028}{{\ttfamily 2009.00028}}].

\bibitem{Lukes-Gerakopoulos:2021ybx}
G.~Lukes-Gerakopoulos and V.~Witzany, \emph{{Non-linear effects in EMRI
  dynamics and their imprints on gravitational waves}},
  \href{https://arxiv.org/abs/2103.06724}{{\ttfamily 2103.06724}}.

\bibitem{Destounis:2021mqv}
K.~Destounis, A.G.~Suvorov and K.D.~Kokkotas, \emph{{Gravitational-wave
  glitches in chaotic extreme-mass-ratio inspirals}},
  \href{https://doi.org/10.1103/PhysRevLett.126.141102}{\emph{Phys. Rev. Lett.}
  {\bfseries 126} (2021) 141102}
  [\href{https://arxiv.org/abs/2103.05643}{{\ttfamily 2103.05643}}].

\bibitem{Lukes-Gerakopoulos:2012qpc}
G.~Lukes-Gerakopoulos, \emph{{The non-integrability of the Zipoy-Voorhees
  metric}}, \href{https://doi.org/10.1103/PhysRevD.86.044013}{\emph{Phys. Rev.
  D} {\bfseries 86} (2012) 044013}
  [\href{https://arxiv.org/abs/1206.0660}{{\ttfamily 1206.0660}}].

\bibitem{Zelenka:2017aqn}
O.~Zelenka and G.~Lukes-Gerakopoulos, \emph{{Chaotic motion in the
  Johannsen-Psaltis spacetime}},  in \emph{{Workshop on Black Holes and Neutron
  Stars}}, 11, 2017 [\href{https://arxiv.org/abs/1711.02442}{{\ttfamily
  1711.02442}}].

\bibitem{Lukes-Gerakopoulos:2017jub}
G.~Lukes-Gerakopoulos and O.~Kop\'a\v{c}ek, \emph{{Recurrence Analysis as a
  tool to study chaotic dynamics of extreme mass ratio inspiral in signal with
  noise}}, \href{https://doi.org/10.1142/S0218271818500104}{\emph{Int. J. Mod.
  Phys. D} {\bfseries 27} (2017) 1850010}
  [\href{https://arxiv.org/abs/1709.08446}{{\ttfamily 1709.08446}}].

\bibitem{Destounis:2021rko}
K.~Destounis and K.D.~Kokkotas, \emph{{Gravitational-wave glitches: Resonant
  islands and frequency jumps in nonintegrable extreme-mass-ratio inspirals}},
  \href{https://doi.org/10.1103/PhysRevD.104.064023}{\emph{Phys. Rev. D}
  {\bfseries 104} (2021) 064023}
  [\href{https://arxiv.org/abs/2108.02782}{{\ttfamily 2108.02782}}].

\bibitem{Mukherjee:2022dju}
S.~Mukherjee, O.~Kopacek and G.~Lukes-Gerakopoulos, \emph{{Resonance crossing
  of a charged body in a magnetized Kerr background: an analogue of extreme
  mass ratio inspiral}},  \href{https://arxiv.org/abs/2206.10302}{{\ttfamily
  2206.10302}}.

\bibitem{schuster_just_2005}
H.G.~Schuster and W.~Just, \emph{Deterministic chaos an introduction},
  Wiley-VCH (2005).

\bibitem{Barausse:2009aa}
E.~Barausse, E.~Racine and A.~Buonanno, \emph{{Hamiltonian of a spinning
  test-particle in curved spacetime}},
  \href{https://doi.org/10.1103/PhysRevD.85.069904}{\emph{Phys. Rev. D}
  {\bfseries 80} (2009) 104025}
  [\href{https://arxiv.org/abs/0907.4745}{{\ttfamily 0907.4745}}].

\bibitem{Steinhoff:2015ksa}
J.~Steinhoff, \emph{{Spin gauge symmetry in the action principle for classical
  relativistic particles}},  \href{https://arxiv.org/abs/1501.04951}{{\ttfamily
  1501.04951}}.

\bibitem{Faye:2006gx}
G.~Faye, L.~Blanchet and A.~Buonanno, \emph{{Higher-order spin effects in the
  dynamics of compact binaries. I. Equations of motion}},
  \href{https://doi.org/10.1103/PhysRevD.74.104033}{\emph{Phys. Rev. D}
  {\bfseries 74} (2006) 104033}
  [\href{https://arxiv.org/abs/gr-qc/0605139}{{\ttfamily gr-qc/0605139}}].

\bibitem{Zhang:2020rxy}
C.~Zhang, W.-B.~Han and S.-C.~Yang, \emph{{Analytical effective one-body
  formalism for extreme-mass-ratio inspirals with eccentric orbits}},
  \href{https://doi.org/10.1088/1572-9494/abfbe4}{\emph{Commun. Theor. Phys.}
  {\bfseries 73} (2021) 085401}
  [\href{https://arxiv.org/abs/2001.06763}{{\ttfamily 2001.06763}}].

\bibitem{Zhang:2021fgy}
C.~Zhang, W.-B.~Han, X.-Y.~Zhong and G.~Wang, \emph{{Geometrized
  effective-one-body formalism for extreme-mass-ratio limits: Generic orbits}},
  \href{https://doi.org/10.1103/PhysRevD.104.024050}{\emph{Phys. Rev. D}
  {\bfseries 104} (2021) 024050}
  [\href{https://arxiv.org/abs/2102.05391}{{\ttfamily 2102.05391}}].

\bibitem{1979GReGr..10...79B}
S.~{Benenti} and M.~{Francaviglia}, \emph{{Remarks on certain separability
  structures and their applications to general relativity}},
  \href{https://doi.org/10.1007/BF00757025}{\emph{General Relativity and
  Gravitation} {\bfseries 10} (1979) 79}.

\bibitem{Cao:2017ndf}
Z.~Cao and W.-B.~Han, \emph{{Waveform model for an eccentric binary black hole
  based on the effective-one-body-numerical-relativity formalism}},
  \href{https://doi.org/10.1103/PhysRevD.96.044028}{\emph{Phys. Rev. D}
  {\bfseries 96} (2017) 044028}
  [\href{https://arxiv.org/abs/1708.00166}{{\ttfamily 1708.00166}}].

\bibitem{Verhaaren:2009md}
C.~Verhaaren and E.W.~Hirschmann, \emph{{Chaotic orbits for spinning particles
  in Schwarzschild spacetime}},
  \href{https://doi.org/10.1103/PhysRevD.81.124034}{\emph{Phys. Rev. D}
  {\bfseries 81} (2010) 124034}
  [\href{https://arxiv.org/abs/0912.0031}{{\ttfamily 0912.0031}}].

\bibitem{lichtenberg_lieberman_1992}
A.J.~Lichtenberg and M.A.~Lieberman, \emph{Regular and chaotic dynamics},
  Springer (1992).

\bibitem{Voglis_1998}
N.~Voglis and C.~Efthymiopoulos, \emph{Angular dynamical spectra. a new method
  for determining frequencies, weak chaos and cantori},
  \href{https://doi.org/10.1088/0305-4470/31/12/015}{\emph{Journal of Physics
  A: Mathematical and General} {\bfseries 31} (1998) 2913}.

\end{thebibliography}\endgroup

\end{document}